# Characterizing Public Debt Cycles: Don't Ignore the Impact of Financial Cycles

Tianbao Zhou[1], Zhixin Liu[1], Yingying Xu[2]*

[1] School of Economics and Management, Beihang University, Beijing, China

[2] School of Humanities and Social Science (School of Public Administration), Beihang University, Beijing, China

*Corresponding author: E-mail: yingxu21@buaa.edu.cn

**Abstract:** Based on the quarterly data from 26 advanced economies (AEs) and 18 emerging market economies (EMs) over the past two decades, this paper estimates the short- and medium-term impacts of financial cycles on the duration and amplitude of public debt cycles. The results indicate that public debt expansions are larger than their contractions in duration and amplitude, aligning with the "deficit bias hypothesis" and being more pronounced in EMs than in AEs. The impacts of various financial cycles are different. Specifically, credit cycles in EMs significantly impact the duration and amplitude of public debt cycles. Notably, short- and medium-term credit booms in EMs shorten the duration of public debt contractions and reduce the amplitude. Fast credit growth in AEs prolongs the duration of public debt expansions and increases the amplitude. However, credit cycles in AEs show no significant impact. For house price cycles, the overall impact is stronger in EMs than in AEs, differing between short- and medium-term cycles. Finally, the impact of equity price cycles is significant in the short term, but not in the medium term. Equity price busts are more likely to prolong the expansion of public debt in EMs while increasing the amplitude of public debt contractions in AEs. Uncovering the impacts of multiple financial cycles on public debt cycles provides implications for better debt policies under different financial conditions.

# 1 Introduction

The Global Financial Crisis (GFC) in 2008 has a profound and far-reaching impact on the economy and public finance of countries worldwide. This is characterized by the decelerating GDP growth, rising borrowing costs and increasing deficits due to government bailouts. One of the most salient issues is that the large accumulation of leverage in financial sectors preceding the crisis, coupled with the burst of the asset price bubble during the crisis, subsequently triggered a prolonged and substantial surge in the public debt level across countries that had already reached to historical highs since the 1870s (Reinhart et al., 2012). The government's debt burden and the risks of debt crisis continue to escalate due to financial instability and sluggish recoveries (Schularick, 2013). For instance, the United States, as the epicenter of the crisis, experienced a considerable shock with its public debt-to-GDP ratio (henceforth, public debt ratio) increasing from around 60% before the GFC to approximately 100% five years later. As an economy with a developed financial sector and high external dependence, Japan was hit hard by the GFC, and its public debt ratio climbed from around 140% before the crisis to approximately 230% today[1]. For China, the public debt ratio rose from about 24%-29% before the GFC to nearly 40% in five years with a continued upward trend. The European Sovereign Debt Crisis in 2010 was one of the consequences of the GFC, sparking discussions on the supervision and management of public debt under the background of the post-GFC era and amid deepening financial developments (Meier, et al., 2021; Stracca, 2014). Therefore, the role of financial market dynamics in the expansion and contraction of public debt is important and complex, which needs further research.

Various perspectives exist regarding the role of financial markets in the economy and public finance. The "irrelevance view" prevalent in the early twentieth century (Modigliani and Miller, 1958) and the "financial accelerator theory" (Bernanke et al., 1999) represent two significantly different views of the relationship between finance and the economy. Financial busts predominantly involve the burst of bubbles in private sector debt (credit) and asset prices. Some studies suggest that private debt and public debt should not be viewed in isolation. The collapse of credit booms often accompanies by the breakdown of business and the banking sector. It will severely damage long-term economic developments and fiscal stability, and lead to a sharp increase in public debt levels as the balance sheets of the private sector may further deteriorate due to fiscal space constraints (Mian and Sufi, 2010; Schularick and Taylor, 2012). In addition, the high level of public debt itself is a factor in economic recovery. Entering a financial bust under a high public debt level may exacerbate the impact on private sector deleveraging, leading to prolonged economic weakness (Bordo and Haubrich, 2010; Checherita-Westphal and Rother, 2012; Jordá, 2016). Besides, fluctuations in asset prices also affect fiscal accounts and public debt levels. Booms and busts in asset prices may affect fiscal instability and primary balance, thereby influencing expansions and contractions of public debt through automatic stabilizers and discretionary fiscal policies (Agnello et al., 2012; Guajardo et al., 2014; Jaeger and Schuknecht, 2007; Tagkalakis, 2011).

---

[1] Public debt-to-GDP ratio data is sourced from BIS and CEIC. Readers can refer to Appendix A for a comprehensive data source for this study.

The differentiation in fiscal policy responses to financial market fluctuations between advanced economies (AEs) and emerging market economies (EMs) holds implications for the distinct features in their respective public debt cycles. For instance, fiscal policies in EMs exhibit pro-cyclicality while those in AEs tend to be counter-cyclical (Cuadra et al., 2010; Talvi and Vegh, 2005). This may be the result of different sensitivity of government spending to international borrowing costs and inconsistent credit information across economies (Dzhambova, 2021; Guérineaua and Léon, 2019). In terms of credit, compared to AEs, credit booms in EMs do not tend to end with financial crises, reflecting the difference in the impact of credit variables and the underlying causes of credit booms between the different economies (Laeven and Valencia, 2013). Unlike AEs which have already established advanced urban infrastructure, many EMs such as China and India, have undergone rapid urbanization and industrialization in recent decades. The increasing prominence of real estate sectors within the national economy, coupled with reinforced interdependencies with other sectors, particularly underscores the growing significance of real estate industries in these economies (Cai et al., 2020). Risks associated with real estate credit pose the potential for substantial spillover effects on other industries, thereby exerting an impact on overall financial stability (Chan et al., 2016; Zhang et al., 2018). Moreover, a large body of literature emphasizes the need to differentiate between short- and long-term debt when discussing the role of public debt in fiscal matters (Angeletos, 2002; Faraglia et al., 2019). All the discussions emphasize that the impact of financial cycles is non-negligible in analyzing the features of public debt cycles.

This study uses the Weibull Accelerated Failure Time (AFT) survival model with shared fragility to investigate the impact of financial cycles on public debt cycles based on the quarterly data from 26 AEs and 18 EMs. Specifically, we investigate the impacts of three financial cycle indicators, i.e., credit cycle, house price cycle, and equity cycles[2], on the duration and amplitude of public debt cycles. The contributions of this paper are three-folds. First, we examine the impact of financial cycles on public debt cycles from two perspectives, i.e., the duration and amplitude, thereby better understanding the role of financial factors in the accumulation and reduction of public debt in terms of time and flow. Second, we discover the relationship between financial cycles and public debt cycles in the short- and medium-term horizons. This provides insights into sustainable debt policies across different time horizons. Finally, we reveal significant differences in the impact of financial cycles on public debt cycles between AEs and EMs.

The average public debt ratio in AEs is considerably higher than that in EMs, and there are also differences in the historical trends between them. Public debt expansions are longer in duration, larger in amplitudes, and have larger slopes than contractions. This feature is more pronounced in EMs, consistent with the "deficit bias hypothesis" which posits that governments tend to increase expenditures during economic downturns to cushion declines, but lack efforts to rebuild fiscal buffers during economic upturns (Debrun et al., 2009; Poghosyan, 2018). Besides, the empirical results verify the varying impacts of financial cycles on public debt cycles. Generally, credit cycles have a significant impact on public debt cycles in EMs, whereas public debt cycles in AEs are mainly affected by the credit growth rate. The impact of house price cycles is more pronounced in EMs. Short- and medium-term house price busts in EMs are associated with prolonged public debt expansions with larger amplitudes. In contrast, medium-term house price booms prolong public debt contractions and increase their amplitudes in AEs and EMs. Furthermore, the impact of short-term equity price busts on prolonging public debt expansions is more pronounced in EMs, whereas the influence of short-term equity price booms associated with greater debt reduction is more important in AEs.

The remainder of this paper is organized as follows. Section 2 is a literature review, Section 3 introduces the methodology and data, Section 4 presents the empirical findings, Section 5 discusses

---

[2] The definition of financial cycles does not reach a consensus and this study opts to represent financial cycles by credit cycles, house price cycles and equity cycles. This approach aligns with other literature (Claessens et al., 2012; Drehmann et al., 2012; Poghosyan, 2018; Yan and Huang, 2020).

the results and Section 6 concludes.

## 2 Literature Review

*2.1 The nexus between public debt cycles and financial cycles*

Fluctuations in credit and asset prices within financial markets primarily affect government finance and public debt through automatic stabilizers and discretionary fiscal policies (Tagkalakis, 2012). In general, the boom in financial markets can significantly improve fiscal balance, alleviate fiscal pressure, and contribute to the consolidation and reduction of public debt. Conversely, financial busts may exacerbate fiscal accounts, thus triggering a surge in public debt and imposing stress on its sustainability. This cyclical pattern is characterized by the rapid expansion of credit and asset prices to the unsustainable level during financial market booms, leading to the imbalance. Subsequently, busts or sharp declines often occur, sometimes evolving into financial crises (Borio, 2014). Public debt, as a primary means to supplement fiscal revenues, also exhibits an expansion-contraction cyclical pattern. Issuing public debt helps to raise more funds for public expenditures. However, excessive expansion of public debt may result in deficits and inflation, further constraining fiscal space. As public debt approaches its limits and requires consolidation, it enters a contraction to alleviate fiscal pressure. The amplification of economic fluctuations by booms and busts in financial cycles can also be observed at different phases of public debt cycles. While excessive booms in financial markets may cause the economy to overheat and hinder debt sustainability (Adarov, 2021), financial busts increase the pressure on fiscal accounts, making it more difficult to reduce debt. Therefore, studying the impact of financial booms and busts on different phases of the public debt cycle is particularly necessary for sustainable public debt.

Numerous studies have confirmed the asymmetric impact of financial cycles on public debt cycles. For instance, Budina et al. (2015) examine the impact during expansions and contractions of public debt cycles and find that the reduction in public debt during financial sector upturns is smaller than the increase during downturns, resulting in the "public debt bias". Poghosyan (2018) contributes further by employing a survival (duration) analysis framework to explore the impact of financial cycles on the duration of public debt cycles, which confirms the "deficit bias hypothesis". Additionally, Poghosyan (2018) finds that financial busts extend the duration of public debt expansions, but the impact of financial cycles on public debt contractions is not significant.

Theoretically, the government needs to adjust fiscal accounts based on asset prices, otherwise, it may lead to procyclical fiscal policies (Eschenbach and Schuknecht, 2002a; Price and Dang, 2011; Tagkalakis, 2012; Yang et al., 2015). Indeed, a temporary windfall revenue from rising asset prices may be perceived as permanent and subsequently accounted for in higher regular expenditures. This makes governments slow to build up fiscal buffers against falling revenues when asset prices eventually fall.

*2.2 Channels through which financial cycles affect public debt cycles*

According to the literature, financial cycles may impact public debt cycles through various direct and indirect channels. For direct impacts, asset price fluctuations directly affect fiscal revenues and public debt levels by altering household and corporate taxes (Eschenbach and Schuknecht, 2002a). For instance, rising asset prices enhance government income from capital gains taxes and value-added taxes on new residences from household and corporate balance sheets, thereby alleviating debt pressures. Conversely, a decline in asset prices leads to a corresponding decrease in such tax revenues, thus requiring more debt issuance.

Indirectly, financial cycles can affect public debt cycles through impacting market liquidity, wealth effect, social security system, inflation, etc. Firstly, during financial booms, credit expansion and rising asset prices stimulate market transactions and increase market liquidity, generating more transaction-related tax revenues, such as turnover taxes and stamp duties (Eschenbach and Schuknecht, 2002a; Morris and Schuknecht, 2007). Although the unit impact of these revenues is small, the large size of the market makes them non-negligible. Secondly, during financial booms, the rise in asset prices enhances consumer confidence and future expectations through the wealth effect, strengthening consumer spending, which indirectly boosts government revenues and alleviates fiscal pressure (Cronin and McQuinn, 2022). This fact aligns with the "permanent income hypothesis" (Friedman, 1957) and is thought to be more pronounced in EMs where a significant proportion of household wealth comes from real estate. Thirdly, asset prices may also impact the viability of public pension reserve funds that support the social security system. A sharp deterioration in asset prices may affect social security spending and related benefits (lowering pension-related expenditures), as well as the level of contributions paid by employers and employees (Tagkalakis, 2011a). Fourth, since tax systems in some countries are not fully inflation-indexed, credit growth may be associated with inflation and/or real exchange rate appreciation (an increase in the relative price of nontradables) and thereby raise revenues (Bénétrix and Lane, 2011). Finally, changes in asset prices generate a "second-round effect" through the real economy and debt service costs. The deterioration (improvement) in private financial balance sheets adversely (positively) affects output levels through decreasing (increasing) investment, employment and consumption, which influences the fiscal account (Eschenbach and Schuknecht, 2002b).

Meanwhile, during financial busts and crises, the prompt fiscal policies adopted by the government can also impact the fiscal account and public debt level. On the one hand, when asset prices collapse or financial institutions face distress, the government may intervene at a cost to save the economy. If through budget subsidies or expenditures, it will directly affect the budget deficit. If through financial transactions such as asset purchases or equity injections, only the public debt ratio is affected. If providing guarantees to the private sector, the government is liable only when required to provide loan guarantees (Tagkalaki, 2011b). On the other hand, if a collapse in asset prices leads to financial instability and a negative feedback loop in economic activity, the government may implement expansionary fiscal measures to avoid the risk of a comprehensive economic downturn, thereby worsening its budget and increasing debt pressure (Reinhart and Rogoff, 2009).

## 3 Methodology and data

*3.1 Dating cycles in short- and medium-term horizons*

This study employs the algorithm introduced by Harding and Pagan (2002), which extends the so-called BB algorithm developed by Bry and Boschan (1971), to identify financial cycles and public debt cycles. Specifically, three kinds of financial cycles are investigated, i.e., credit cycles, house price cycles and equity price cycles. Firstly, the turning points of the public debt-to-GDP ratio, the private sector credit-to-GDP ratio, the real house price index and the composite stock price index are identified corresponding to the local maxima and minima within a specified interval. For quarterly sequences, this interval consists of the two quarters before and after. Specifically, if there is a local maximum point $Y_t$ at time $t$, then we have

$$(Y_t - Y_{t \pm k}) > 0, k = 2 \tag{1}$$

Likewise, if there is a local minimum point $Y_t$ at time $t$, then we have

$$(Y_t - Y_{t \pm k}) < 0, k = 2 \tag{2}$$

Once all the turning points are identified, we proceed to exclude which do not meet the censoring rules. The local maxima points that pass the censoring rules are referred to as peaks, and the minima points are referred to as troughs. Three censoring rules are employed. First, peaks and troughs must alternate. If multiple consecutive maxima occur, only the highest is retained as the peak; and only the lowest is retained as the trough for consecutive minima. Second, the value of a peak must be higher than the previous trough and the value of a trough must be lower than the previous peak. Following Claessens et al. (2012) and Poghosyan (2018), we define the phase from a peak to a trough as a contraction (bust) of the public debt (financial) cycle, and from a trough to a peak as an expansion (boom) of the public debt (financial) cycle[3]. A complete cycle consists of one expansion (boom) and one contraction (bust) (regardless of their sequential order). Third, for short- and medium-term cycles, each expansion (boom) or contraction (bust) must be at least two and four quarters, respectively. Therefore, a complete short- or medium-term cycle must be at least five or nine quarters.

*3.2 Parametric survival model*

We employ the survival analysis model, also known as the duration analysis model, to examine the impact of financial cycles on the duration of public debt cycles. The survival analysis model is adept at analyzing the exogenous factors that make the duration longer or shorter. Since each country may experience multiple cycles, the Weibull Accelerated Failure Time (AFT) model with shared frailty is used to investigate the impact of financial cycles on the duration of public debt cycles considering the unobserved heterogeneity of countries (Clayton, 1978; Hougaard, 1986). The Weibull AFT model is as follows:

$$\ln T = \boldsymbol{X'\beta} + \mu \qquad (3)$$

where $T$ is the duration of public debt expansions and contractions, $\boldsymbol{X'}$ is the matrix of exogenous variables containing three sets of dummy variables (indicating the association between financial cycles and public debt cycles), credit growth rate, house price growth rate and control variables. These variables are described in detail in the next subsection. $\boldsymbol{\beta}$ is the coefficient matrix of the exogenous variables, and $\mu$ is the residual term following the Weibull distribution. We then have the survival function $S(t) = P(T \geq t)$, the failure function $F(t) = 1 - S(t)$, and its density function $f(t) = F'(t)$. Thereby, the hazard function is $\lambda(t) = f(t)/S(t)$. In the Weibull AFT model with shared frailty, the hazard function for the $j^{th}$ expansion or contraction of country $i$ at survival time $t$ is

$$h_{ij}(t_{ij}, \boldsymbol{X}_{ij}, \alpha_i) = \alpha_i p t_{ij}^{p-1} exp(\boldsymbol{X'}_{ij}\widetilde{\boldsymbol{\beta}}) \tag{4}$$

where $\widetilde{\boldsymbol{\beta}} = -\boldsymbol{\beta}$. The shared frailty of individuals within a sample is represented by $\alpha_i$, which satisfies $\alpha_i > 0$, $E(\alpha_i) = 1$, and $Var(\alpha_i) = \theta$, and is independent of the exogenous variables. If $\theta = 0$, then there is no significant shared frailty. Otherwise, for each sample in this study, i.e., AEs and EMs, $\alpha_i$ is different for distinct country, but all follow an inverse Gaussian distribution.

---

[3] The use of expansion and boom and contraction and bust is to avoid confusing the two cycles, but has the same definition.

*3.3 Data*

The study incorporates quarterly data from 44 countries, encompassing 26 AEs and 18 EMs. Limited by data availability, the starting date for each country differs. However, all data ends at 2022Q4. Data for AEs mainly starts from the 1990s, while data for EMs mostly begins around 2000. This study uses three categories of data, i.e., the public debt cycle, the financial cycle, and macroeconomic variables. We employ the public debt-to-GDP ratio, i.e., public debt ratio, as a proxy for the public debt cycle. For the financial cycle, we use the private sector credit-to-GDP ratio, i.e., credit ratio, the real house price index and the composite stock price index to measure the credit cycle[4], house price cycle and equity price cycle[5], respectively. We also incorporate six macroeconomic variables as control variables for robustness tests. All variables are seasonally-adjusted quarterly data (at constant prices when necessary). For brevity, the detailed information of all variables is provided in Appendix A.

*3.4 Cycles association*

After identifying cycles, we define six dummy variables indicating the association between financial cycles and public debt cycles. In the short-term cycle, when there is a credit peak (trough) within one quarter before and after a public debt trough (peak), we define that the public debt expansion (contraction) is associated with the credit bust (boom). Specifically,

---

[4] Credit is the key channel connecting savings and investment, reflecting supply and demand dynamics. The use of credit ratios better captures the relative scale of credit and the level of financial risk under prevailing economic conditions (Gaiotti, 2013; Poghosyan, 2018).

[5] The real estate price index and composite stock price index represent weighted aggregates of representative price categories reflecting the overall price changes and trends within these two markets.

$$D_t^{credit\ bust} = \begin{cases} 1, \text{if a credit peak appears in } [\,t-1, t+1\,], \text{where } t \text{ is the date of a public debt trough} \\ 0, \text{otherwise} \end{cases} \quad (5)$$

$$D_t^{credit\ boom} = \begin{cases} 1, if \text{ a credit trough appears in } [\,t-1, t+1\,] \text{where } t \text{ is the date of a public debt peak} \\ 0, \text{otherwise} \end{cases} \quad (6)$$

We define the dummy variables for the house price cycle and equity price cycle in the same way as $D_t^{house\ price\ bust}$, $D_t^{house\ price\ boom}$, $D_t^{equity\ price\ bust}$, $D_t^{equity\ price\ boom}$. For the medium-term cycle, we change the interval to two quarters, i.e., $[\,t-2, t+2\,]$, leaving other definitions unchanged. Practically, each expansion (contraction) of public debt may be associated with one or more credit, house price and equity price busts (booms), or none of them. Note that the association between financial cycles and public debt cycles is only the fact we observe but not causality.

*3.5 Descriptive analysis of data*

Fig. 1 presents the trend and average of the public debt ratios over the past two decades for 26 AEs and 18 EMs. It shows that the level of public debt ratios in AEs is significantly higher than in EMs, and from the early 21st century to the GFC, the average public debt ratio in EMs (AEs) had dropped around 40% (50%). This period coincides with the robust global growth after the dot-com bubble, characterized by stable economic expansion and modest fiscal pressure. After the GFC till the outbreak of the COVID-19 pandemic in 2020, The average public debt ratio in AEs surged from 50% to nearly 100%, whereas the average public debt ratio in EMs moderately increased from around 40% to 60%. In the post-COVID era, AEs showed significant drops in public debt ratios without rebounding, showing the efficacy of debt reduction by the governments. However, the average public debt ratio in EMs resumed the upward trend to approximately 56% and followed a brief retreat. These findings underscore the pronounced disparities in the level and historical trend of public debt ratios between AEs and EMs, warranting a distinct analysis in further parametric examination.

**(Insert Fig. 1 here)**

This study identifies public debt cycles and financial cycles in 44 economies. Table 1 reports their duration, amplitude, and slope, which measure the persistence, cumulative effect, and degree of trend change of the corresponding cycle. In short-term (medium-term) cycles, AEs exhibit 153 (84) public debt expansions and 136 (69) public debt contractions, whereas EMs show 72 (33) public debt expansions and 67 (29) public debt contractions. Panel A presents the results for public debt expansions. In the medium term, the duration of public debt expansions in EMs is significantly longer than that in AEs, with relatively small differences in the short term. Regardless of the time horizon, EMs consistently show higher amplitude and slope in expansions and contractions than AEs and also display greater variance in these two metrics. Panel B presents the results for public debt contractions. Although the duration of public debt contractions is slightly larger in EMs than that in AEs in the medium term, the duration, amplitude and slope of the public debt contractions are generally similar in both groups, in the short and medium term. Moreover, expansions are significantly longer, more substantial and steeper than contractions, with this feature being more pronounced in EMs. For instance, in the medium term, the average contraction amplitude for AEs and EMs is around -14%, whereas the average expansion amplitude is 51.55% and 106.64% for AEs and EMs, respectively. These findings are consistent with the "deficit bias hypothesis" (Debrun et al., 2009; Poghosyan, 2018), suggesting that governments tend to increase expenditures during economic downturns (or recessions) to cushion declining growth but lack efforts to rebuild fiscal buffers during economic upturns.

**(Insert Table 1 here)**

Fig. 2 shows the average duration of public debt expansions associated with financial busts. The results of AEs indicate that in the short (medium) term, the average duration of public debt expansions associated without any financial busts is approximately 6.7 (12) quarters. The average duration of public debt expansions associated with financial busts ranges from 8.7 to 10 (15.7 to 19) quarters in the short (medium) term. For EMs, in the short (medium) term, public debt expansions last about 6.3 (13.2) quarters in the absence of any financial busts. Public debt expansions associated with financial busts have an average duration ranging from 8 to 14.9 (19 to 31.8) quarters. Generally, public debt expansions associated with financial busts are longer lasting than those without such busts, which is particularly pronounced in EMs. It suggests that, in general, financial busts may prolong the expansion of public debt.

**(Insert Fig. 2 here)**

Fig. 3 shows the average duration of public debt contractions associated with financial booms. The results of AEs indicate that, in the short (medium) term, the average duration of public debt contractions associated without any financial booms is approximately 5.9 (11) quarters, whereas the value ranges from 7.1 to 10.1 (14.3 to 16) quarters when associated with financial booms. For EMs in the short (medium) term, public debt contractions associated without any financial booms last about 6.7 (13.4) quarters. The average duration of public debt contractions associated with financial booms varies. Besides, public debt contractions associated with credit booms last for three to four quarters. In the medium term, public debt contractions associated with house price booms and equity price booms are longer than those without such booms. In the short term, only public debt contractions associated with equity price booms are longer.

In summary, the average duration of public debt expansions associated with financial busts is longer than those without busts, whereas the average duration of public debt contractions associated with financial booms varies across economies and time horizons.

**(Insert Fig. 3 here)**

## 4 Empirical results

*4.1 The impact of financial cycles on the duration of public debt cycles*

In this section, we employ the Weibull AFT model with shared frailty to examine the impact of financial cycles on the duration of public debt cycles. Specifically, we examine whether financial busts (booms) prolong or shorten the duration of public debt expansions (contractions). Table 2 presents the results of short-term public debt expansions in EMs. M1 is the benchmark model, controlling only for the shared frailty. M2 to M4 examine the impact of credit busts, house price busts and equity price busts on the duration of public debt expansions, respectively. The results show that the coefficients of house price bust (0.6155 in M3) and equity price bust (0.6267 in M4) are significantly positive. It suggests that house price busts prolong the duration of public debt expansions by 1.85 ($e^{0.6155}$) times lengthier, and equity price busts prolong public debt expansions by 1.87 ($e^{0.6267}$) times longer compared to those without such busts. Credit bust in M2 (-0.1247) shows no significant impact on the duration of public debt expansions. M5 includes the three dummy variables, and M6-M8 further consider credit growth rate and house price growth rate. The results for M5-M8 show that the coefficients of house price bust and equity price bust remain significantly positive, whereas the coefficient of credit bust remains statistically insignificant. This is consistent with the results from M2-M4, indicating that there is no multi-collinearity among variables and confirming the reliability of the model estimates. All Weibull parameters $p$ are significantly greater than one, indicating a positive time-dependence in public debt expansions. It implies that as the expansion lasts, the likelihood of ending itself increases.

**(Insert Table 2 here)**

Based on Table 2, we use M5 and M8 to proceed the following empirical analysis for brevity. Specifically, the complete results for all models are provided in Appendix B. Panel A in Table 3 presents the results of the impact of financial busts (credit busts, house price busts and equity price busts) on the duration of public debt expansions. First, in the short term, the coefficient of equity price bust in EMs is significantly positive (0.5593 and 0.4362). However, the coefficient of equity price bust in AEs is notably smaller than that of EMs in the short term. It suggests that, in the short term, equity price busts prolong the expansion of public debt in both economies, but it is more pronounced in EMs[6]. Second, the coefficient of house price bust in EMs is significant in the short and medium term, indicating that house price busts significantly prolong the expansion of public debt. However, the impact of house price busts in AEs is insignificant. Third, for credit busts, only the coefficient in EMs in the medium term is significantly positive, which means that credit busts prolong the expansion of public debt in EMs in the medium term. Finally, in AEs, the coefficient of credit growth rate is significantly positive. Meanwhile, the coefficient of house price growth rate is significantly negative in the short term. This suggests that a higher credit growth rate and a lower house price growth rate two quarters before a public debt expansion lead to a longer subsequent expansion.

Panel B in Table 3 presents the results of the impact of financial booms (credit booms, house price booms and equity price booms) on the duration of public debt contractions. The house price growth rate is dropped from Panel B because this variable causes significant multi-collinearity in the medium-term contraction regressions. First, in the short term, the coefficient of equity price boom is significant in AEs and EMs. It shows that equity price booms prolong public debt contractions in the short term and this impact is similar in the two economies[7]. Second, in the medium term, the coefficient of house price boom is significant in AEs and EMs, suggesting that house price booms prolong public debt contractions, and this impact is similar in two groups of countries[8]. However, short-term house price booms in EMs slightly shorten public debt contractions, an effect that is absent in AEs. Third, in the short and medium term, credit booms in EMs shorten public debt contractions, whereas in AEs, short-term credit booms mildly prolong public debt contractions.

---

[6] When combining the samples of both types of economies and introducing a dummy variable representing EMs, the coefficient of the interaction term between this dummy variable and equity price bust is significantly greater than zero, suggesting that the effect of equity price busts on public debt expansions is more pronounced in EMs. See details in Appendix C.

In addition, all Weibull parameters $p$ in Table 3 are significantly greater than one. This implies a positive time-dependence in both the expansion and contraction of public debt, regardless of the economies' type and time horizons. Each expansion and contraction is more likely to end over time, which is evidence of the existence of the public debt cycle. The alternating pattern of public debt expansions and contractions over time creates the recurrent public debt cycles.

**(Insert Table 3 here)**

*4.2 The impact of financial cycles on the amplitude of public debt cycles*

Financial cycles can also impact the amplitude of public debt expansions or contractions ($Amplitude_{t_1} = \frac{PD_{t_1} - PD_{t_0}}{PD_{t_0}}$, $PD_t$ is the public debt ratio at time $t$). This study proceeds to employ panel regressions with the country's fixed effect to examine the impact of financial busts (booms) on the amplitude of public debt expansions (contractions). Table 4 summarizes the main results, and see Appendix B for the complete results for all regressions. Panel A shows the impact of financial busts on the amplitude of public debt expansions. First, the coefficient of medium-term credit bust in EMs is significant, meaning that credit busts increase the amplitude of public debt expansions by 58% to 67%. Second, the coefficient of house price bust is significant only in the short term, but the sign of the coefficient is opposite in AEs and EMs. House price busts mildly decrease the amplitude of public debt expansions in AEs by approximately 19.48%, whereas increasing it by about 27%-29% in EMs. Third, for the short-term equity price bust, the coefficient is positive, although with a weak significance for AEs. Finally, the coefficient of credit growth rate is significantly positive in AEs. The coefficient of house price growth is significantly negative in the short term, indicating that a higher credit growth rate and a lower house price growth rate two quarters before a public debt expansion lead to a larger amplitude of the subsequent expansion.

---

[7] When combining the samples of both types of economies and introducing a dummy variable representing EMs, the coefficient of the interaction term between this dummy variable and equity price boom is not significant, suggesting that the effect of equity price booms on public debt contractions is similar in AEs and EMs. See details in Appendix C.

[8] When combining the samples of both types of economies and introducing a dummy variable representing EMs, the coefficient of the interaction term between this dummy variable and house price boom is not significant, suggesting that the effect of house price booms on public debt contractions is similar in AEs and EMs. See details in Appendix C.

Panel B presents the results on the impact of financial booms on the amplitude of public debt contractions. First, credit booms have a significant inhibitory effect on public debt contractions in EMs, which is not conducive to public debt reduction. Specifically, credit booms decrease the amplitude of public debt contractions by 6%, and this value is approximately 15% in the medium term. However, credit booms do not have a significant impact on the amplitude of public debt contractions in AEs. Second, house price booms increase the amplitude of public debt contractions in EMs in the medium term. Finally, equity price booms increase the amplitude of public debt contractions in AEs in the short term.

**(Insert Table 4 here)**

*4.3 Robustness tests for the impact of financial cycles on public debt cycles*

This study adopts a variety of methods to test the robustness of the model and the empirical results. Firstly, according to the literature, we add six macroeconomic indicators as control variables, i.e., real GDP growth rate, money supply growth rate, inflation rate, real effective exchange rate (REER) growth rate, account balance, and crude oil price growth rate. In the robustness test for the duration of public debt cycles, each macroeconomic variable is added to the model one by one, and the results show that the impact of financial cycles on public debt cycles is robust. For brevity, we present the complete results in Appendix D. Second, considering the possible multi-collinearity of the six macroeconomic variables, the first three principal components of the six variables (the variance contribution rate is about 70%) are further extracted and added into the model as control variables. The results in Table 5 show that the inclusion of the first three principal components of macroeconomic variables does not essentially change the impact of financial cycles on the duration of public debt cycles, and the feature of Weibull parameter $p$ in the models maintains unchanged. Finally, this study considers the impact of orthogonalized financial cycles on public debt cycles to eliminate the possible impact of credit growth rate and house price growth rate on financial cycles. Again, the results in Appendix D show the robustness of our empirical results. Similar robustness tests are conducted for investigating the amplitude of public debt cycles. All results consistently show that accounting for economic conditions and time-varying factors across economies does not change the impact of financial cycles on public debt cycles.

**(Insert Table 5 here)**

## 5 Discussions

*5.1 The impact of credit cycles on public debt cycles*

The results presented in this study show that the impact of credit cycles on public debt cycles is not significant in AEs, but highly significant in EMs. For AEs, the duration and amplitude of the subsequent public debt expansion are likely to be longer and larger if there is a higher credit growth rate two quarters before the start of the expansion. However, the impact of credit growth rate on public debt contractions is not significant in AEs. The emergence of public debt expansions is always associated with economic downturns or crises in AEs, and the rapid growth of private sector debt and excessive leverage preceding such periods may drag down the subsequent economic recovery, resulting in higher fiscal costs (Law and Singh, 2014; Samargandi et al., 2015) and additional public debt for governments (Bernardini and Forni, 2017). In contrast, neither credit cycles nor the credit growth rate in AEs affect public debt contractions significantly, possibly because the reduction of public debt in AEs is primarily driven by factors other than credit, such as house prices, equity prices, tax revenues and fiscal expenditures. Generally, the impact of credit variables in AEs is mainly reflected in the growth rate, whereas it manifests more as cyclical dynamics in EMs, thus emphasizing the importance of selecting and constructing credit indicators based on economy type in future research (Geršl and Jašová, 2018).

In EMs, the impact of credit cycles on public debt cycles differs between the short and medium term and between expansions and contractions. Short-term credit busts do not significantly affect the duration and amplitude of public debt expansions. However, medium-term credit busts are associated with prolonged public debt expansions with greater debt accumulation. Short- and medium-term credit booms are associated with shorter public debt contractions and less debt reduction. In EMs, despite short-term credit busts do not significantly impact public debt expansions, in a longer horizon, medium-term credit busts may negatively affect total demand and asset prices through increasing risk premiums, disrupting credit supply to bank-dependent borrowers, and reducing consumption and investment (Alfaro et al., 2021; Cingano et al., 2016). This has a detrimental effect on economic growth and fiscal accounts in EMs, as their developments mainly rely on private sector investment, further exacerbating public debt expansions aimed at encouraging investment and stimulating the economy. Loose monetary policy is also identified as the most important factor in the short- and long-term domestic credit expansions in EMs (Gozgor, 2014; Guo and Stepanyan, 2011). The implementation of loose monetary policy is typically accompanied by banks purchasing bonds from the public sector, accompanied by an increase in public debt issuance. Additionally, credit booms in EMs are less likely to be associated with systemic banking crises, as the booms are usually caused by financial deepening rather than the accumulation of financial risks (Meng and Gonzalez, 2016). Therefore, governments in EMs may not be inclined to sacrifice growth opportunities and incur high opportunity costs by curbing credit booms. Moreover, credit booms in EMs, low-to-middle-income countries, and countries with weak exchange rate flexibility are often accompanied by substantial capital inflows (Dell'Ariccia et al., 2016; Zehri and Madjd-Sadjadi, 2023). This may result in increasingly fragile balance sheets of banks and corporations, soaring asset prices, strong domestic demand and expanding deficits (Elekdag and Wu, 2013). Governments may issue more public debt to raise revenues, thereby reducing the duration and amplitude of public debt contractions. Generally, rapid credit growth, credit busts and credit booms are found to either exacerbate public debt expansions or curb public debt contractions. Therefore, implementing strict macro-prudential policies to ensure the stability of credit levels is crucial for sustainable public debt development.

*5.2 The impact of house price cycles on public debt cycles*

For EMs, short- and medium-term house price busts are associated with prolonged public debt expansions with greater amplitudes. However, while short-term house price booms are associated with shortened public debt contractions, medium-term house price booms are associated with prolonged public debt contractions with larger amplitudes. This suggests that, in EMs, the impact of house price cycles on public debt cycles differs over time horizons. By contrast, the impact of house price cycles in AEs is much weaker than that in EMs. Specifically, in AEs, only medium-term house price booms prolong public debt contractions, whereas their impact in the short term is not significant.

For AEs and EMs, the different importance of the housing market in economic growth and fiscal revenues may be one reason for the marked difference in the impact of house price cycles on public debt cycles. For EMs, a larger proportion of household assets is occupied by housing, with a higher housing-to-income ratio (Lu et al., 2020). For instance, in 2016, housing assets accounted for approximately 54% of the net worth of Chinese households, whereas in some European AEs and the United States, this figure was only 20%-30% (Li and Zhang, 2021). In addition, the real impact of the real estate sector on the economy in EMs may be greater than the contribution to total value added, with various financial channels connecting upstream and downstream industries, influencing economic growth and public debt (Chan et al., 2016). These factors result in more significant economic shocks from house price fluctuations in EMs compared to AEs. Notably, the industrialization process in EMs over recent decades has greatly benefited from the expansion and boom in the real estate sector. Related taxes, such as land-transfer fees, property taxes and turnover taxes, play a crucial role in fiscal balance and local economic development (Gibb and Hoesli, 2003; Zhang et al., 2016). Therefore, government deficits are more likely to worsen due to reduced fiscal revenues from declining house prices. It may escalate fiscal expenditures aimed at stabilizing markets and result in decreased household consumption due to asset depreciation, further intensifying public debt expansion. However, studies of AEs such as the United States suggest that property tax revenues do not necessarily decline during house price busts. This can be attributed to the lag in converting falling house prices into assessed values. Local policymakers counter the decline in property tax revenues by raising actual tax rates while maintaining the tax base unchanged. Furthermore, during periods of higher house prices, policymakers tend to lower effective tax rates to balance the increased property tax revenues, thus ensuring fiscal stability (Goodman, 2018; Lutz et al., 2011). Besides, AEs possess larger rental markets, where houses serve more as necessities than investments. Therefore, the impact of house price busts on household consumption and asset value is less pronounced in AEs.

On the other side, medium-term house price booms in AEs and EMs significantly stimulate consumption through increased government tax revenues from the real estate sector and strong wealth effects. According to the "permanent income hypothesis" (Friedman, 1957) and the life-cycle theory (Ando and Modigliani, 1963), consumers' spending responds only to permanent changes in wealth. While financial wealth, e.g., securities, is vulnerable to temporary shocks that cause short-term fluctuations, housing assets often experience more persistent shocks and affect long-term household wealth. Meanwhile, the housing wealth effect is more pronounced in EMs (Kishor, 2007; Singh, 2022). Therefore, medium-term house price booms manifest in alleviating fiscal pressure and enhancing public debt contractions through the consumption channel. Besides, rising house prices also drive the development of ancillary facilities and investment in upstream and downstream industries. Compared with AEs, infrastructure construction in EMs stimulates local economic growth (Han, et al., 2021) and influences housing prices, fiscal balance and debt issuance.

Notably, short-term house price booms slightly shorten debt contractions in EMs. One possible reason is that short-term house price booms increase market volatility and people may not immediately expand consumption in response to short-term appreciation of housing assets, i.e., the "permanent income hypothesis" and the life-cycle theory. Instead, due to rising housing costs and market uncertainty, people may increase precautionary savings and reduce expenditures. Research on China indicates that rising house prices also have a crowding-out effect on other sectors, leading to capital misallocation, decline in manufacturing and pressure on R&D (Lu et al., 2019; Wu et al., 2020). The rising land cost also reduces the competitiveness of large cities (Li and Zhang, 2021), making the government in EMs with unbalanced regional development incur certain fiscal costs to adjust for the impact of rising housing prices, resulting in a decrease in fiscal revenues and weakening the amplitude of public debt contractions in the short term.

*5.3 The impact of equity price cycles on public debt cycles*

In the short term, equity price cycles have a significant impact on public debt cycles, but not in the medium term. Short-term equity price busts are associated with prolonged public debt expansions with greater amplitudes. Short-term equity price booms are associated with more prolonged public debt contractions with larger debt reductions. However, the impact of short-term equity price cycles varies between AEs and EMs. The impact of equity price busts on public debt expansions is stronger in EMs, whereas the impact of equity price booms on public debt contractions is more pronounced in AEs. It appears that AEs have stronger fiscal stability in the face of stock market declines and better utilize equity price booms to strengthen debt reduction. In contrast, EMs show increased debt pressure during stock declines and lack the ability to undertake more robust debt cut-downs during equity price booms.

The short-term fluctuations in equity prices are closely related to output growth. The short-term equity price cycle reflects investors' future expectations, consumers' willingness, and enthusiasm of businesses to invest (McMillan, 2020; Yilanci et al., 2021)[9]. When equity prices rise, revenues from transaction taxes, income taxes, and commodity taxes increase, enabling a more persistent and substantial public debt reduction. Conversely, these taxes decrease when equity prices fall, accompanied by a decline in domestic investment and consumer enthusiasm. In response, the government may issue more public debt to compensate for reduced income and stimulate demand. The limited impact of medium-term equity price cycles on public debt cycles suggests equity prices are not suitable as a long-term debt management indicator. This may be attributed to temporary shocks that dominate changes in household financial wealth, affecting consumer spending and overall demand only in the short term (Kishor, 2007). Variables such as house prices, output and credit, which have more sustained impacts on household wealth, may be more suitable for analyzing public debt issuance over longer horizons.

One possible explanation for the differential impact of short-term equity price cycles between AEs and EMs is the stock wealth effect. AEs exhibit a stronger stock wealth effect with a higher degree of marketization (Carroll et al., 2011). EMs are characterized by lower market liquidity, which may incur higher transaction costs, resulting in smaller impacts on consumption (Sonje et al., 2014). Therefore, the rise in equity prices in AEs can more effectively stimulate household income and boost consumption, which in turn increases overall demand and raises the government's tax revenues for public debt reduction. Simultaneously, the increase in GDP due to rising equity prices also enlarges the denominator of the public debt ratio, leading to a larger public debt contraction in AEs. However, the prolonged impact of equity price busts on public debt expansions suggests that EMs should exert more effort to maintain stock market stability in the future.

## 6 Conclusion

This study examines the impact of financial cycles on public debt cycles in 44 economies. Based on the historical public debt ratios, we find that public debt expansions are longer in duration, larger in amplitude, and have larger slopes than contractions. This feature is consistent with the "deficit bias hypothesis" (Debrun et al., 2009; Poghosyan, 2018) and is more pronounced in EMs. The empirical results show that the impact on public debt cycles brought by financial cycles differs between the short and medium term and between AEs and EMs. First, for credit cycles, the impact on public debt cycles is significant in EMs, but not in AEs. It is the credit growth rate rather than credit cycles that affect public debt cycles in AEs. Second, the impact of house price cycles on public debt cycles is more pronounced in EMs than in AEs. Specifically, in EMs, short- and medium-term house price busts are associated with prolonged public debt expansions with larger amplitudes, whereas short-term (medium-term) house price booms are associated with shortened (prolonged) public debt contractions. Besides, house price booms are only associated with prolonged public debt contractions in the medium term in AEs. Third, the impact of equity price cycles on public debt cycles is significant in the short term, but not in the medium term. On the one hand, short-term equity price busts are associated with prolonged public debt expansions with larger amplitudes. On the other hand, short-term equity price booms are associated with longer public debt contractions with more debt reduction. Notably, equity price busts have a stronger effect on prolonging the duration of public debt expansions in EMs, whereas the influence of short-term equity price booms associated with greater debt reduction is more important in AEs.

---

[9] McMillan (2020) and Yilanci et al. (2021) find that equity prices have the strongest predictive effect on output growth within approximately four quarters, aligning with the short-term public debt expansion and contraction discussed in this paper.

The conclusions of this study provide several policy implications. First, since rapid growth, booms and busts of credit could either exacerbate public debt expansions or curb public debt contractions, controlling fluctuations in credit levels is crucial for maintaining the stability of public debt in AEs and EMs over any time horizon. Second, governments of EMs should implement more effective policies to support the sustainable development of the real estate sector and build robust fiscal buffers and avoid excessive conversion into fiscal expenditures during house price booms. This helps to better cushion the sharp increase in public debt caused by falling house prices. Besides, governments of EMs should reduce their dependence on the real estate industry taxes by advancing the transformation of land finance and diversifying fiscal income through urban management, innovation and entrepreneurship. Third, governments should emphasize the importance of the stock market in short-term debt management. By promoting the sustainable development of the stock market, public debt can be reduced more effectively during debt contractions. EMs also need to accelerate the establishment of comprehensive stock market mechanisms and introduce policies and regulations to enhance market stability, thereby reducing the prolonged expansions of public debt caused by equity price busts. This study is limited by the sample of credit (private sector debt) data for EMs. In comparison to house price cycles and equity price cycles, we identify a relatively small quantity of associations between credit cycles and public debt cycles, which may have affected the reliability of the findings in this part. As more data become available, future research can be further deepened in the selection of credit indicators, the identification of credit cycles, and the dynamic relationship between credit cycles and public debt cycles.

## Data Availability Statement

The data are available from BIS, CEIC database, OECD, and IMF.

## Declare of interest

The authors declare no conflict of interest.

## Acknowledgment

This work was supported by the Key Program of National Natural Science Foundation of China [grant number 72033001]; the National Natural Science Foundation of China [grant number 72203019].

# Figures and tables

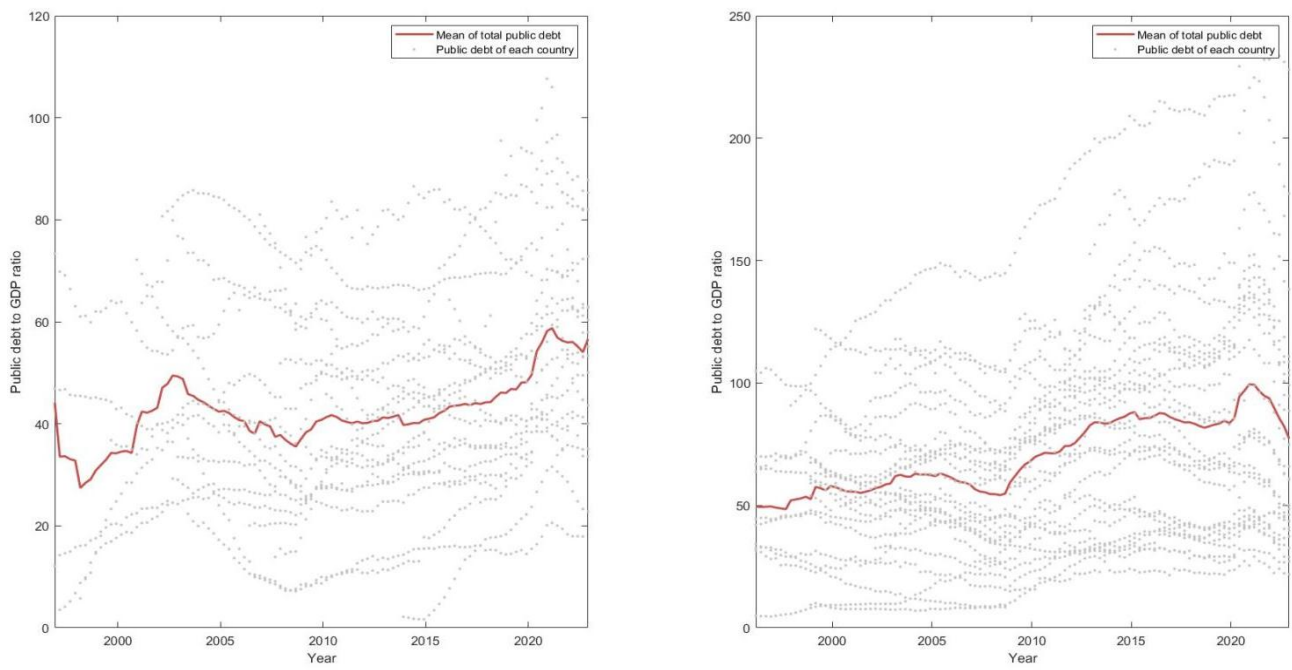

**Fig. 1. Historical public debt to GDP ratios in EMs (left) and AEs (right).**

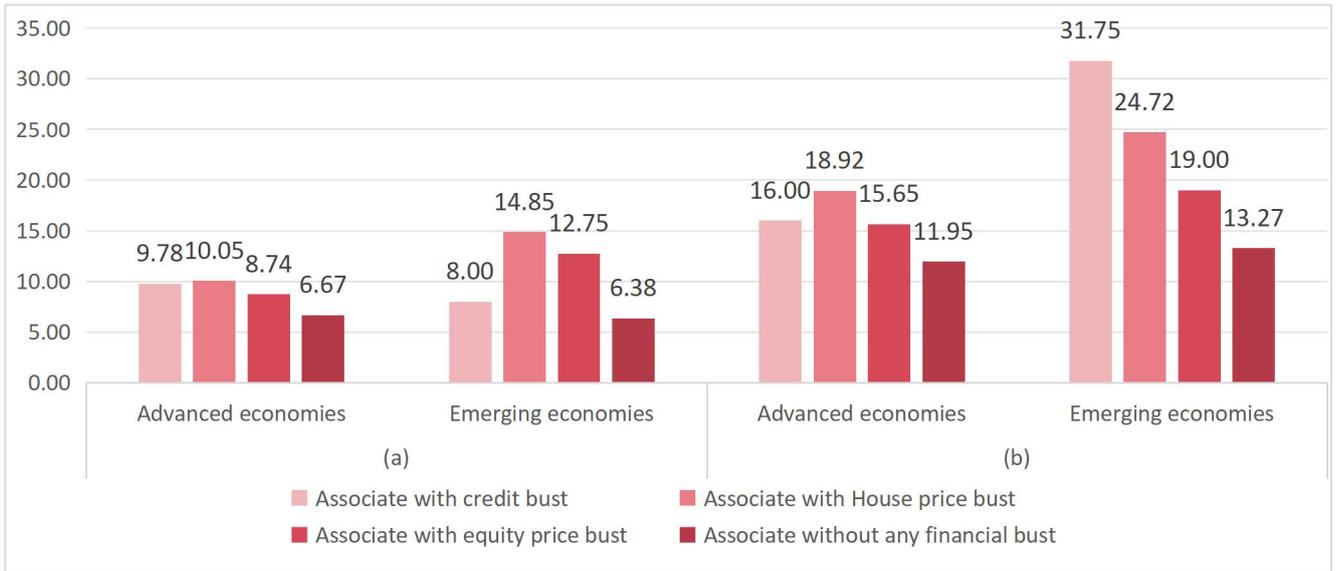

**Fig. 2. Duration of short-term (a) and medium-term (b) public debt expansions associated with financial busts.**

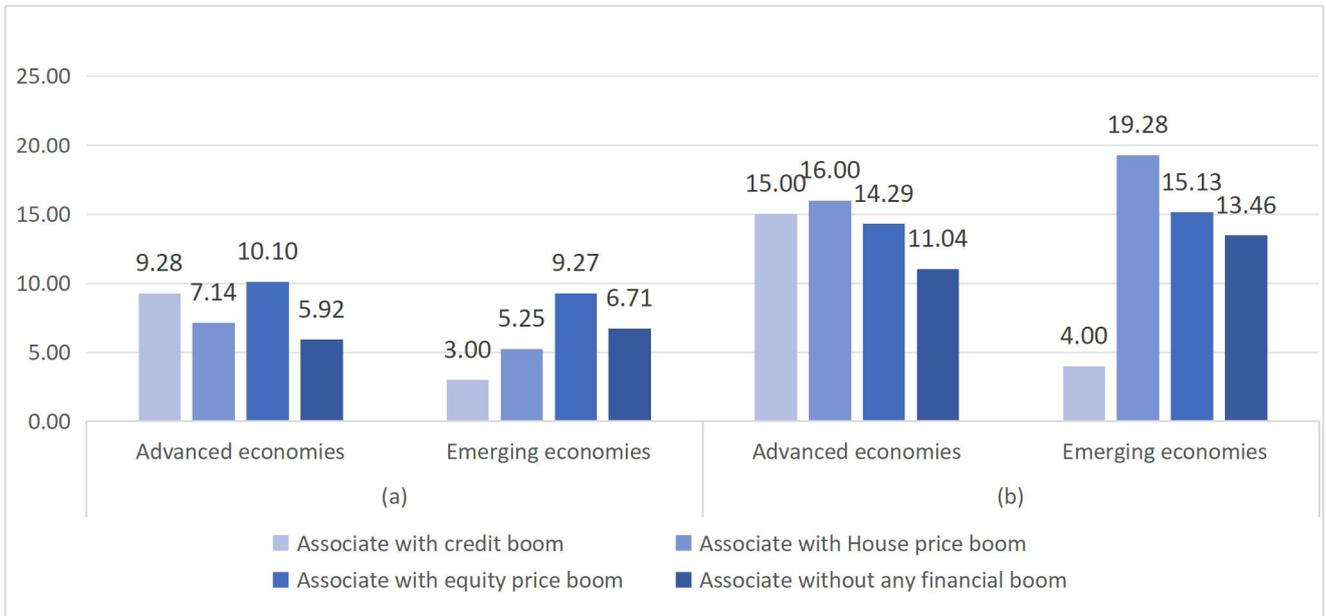

**Fig. 3.** Duration of short-term (a) and medium-term (b) public debt contractions associated with financial booms.

Table 1. Feature of public debt cycles.

| Panel A: Statistics of public debt expansions | | | | | | | |
|---|---|---|---|---|---|---|---|
| | Number of events | Duration (quarters) | Amplitude (% of GDP) | Slope (% of GDP) | Number of associations with credit busts | Number of associations with house price busts | Number of associations with equity price busts |
| **Short-term** | | | | | | | |
| AE | 153 | 7.9085 (6.8320) | 29.4963 (55.9137) | 2.9446 (2.8418) | 9 | 22 | 54 |
| EM | 72 | 9.0417 (9.0155) | 52.5916 (214.2727) | 3.8907 (8.5499) | 4 | 13 | 24 |
| Total | 225 | 8.2711 (7.5971) | 36.8868 (129.5791) | 3.2473 (5.3709) | 13 | 35 | 78 |
| **Medium-term** | | | | | | | |
| AE | 84 | 13.9762 (10.3155) | 51.5526 (77.0525) | 3.1325 (2.7890) | 6 | 14 | 26 |
| EM | 33 | 17.3030 (11.9778) | 106.6396 (311.1243) | 5.0799 (12.2345) | 4 | 7 | 9 |
| Total | 117 | 14.9145 (10.8616) | 67.0899 (177.6819) | 3.6812 (6.0915) | 10 | 21 | 35 |
| **Panel B: Statistics of public debt contractions** | | | | | | | |
| | Number of events | Duration (quarters) | Amplitude (% of GDP) | Slope (% of GDP) | Number of associations with credit booms | Number of associations with house price booms | Number of associations with equity price booms |
| **Short-term** | | | | | | | |
| AE | 136 | 6.9779 (6.7225) | -9.7533 (10.5045) | -1.5061 (1.1574) | 7 | 15 | 26 |
| EM | 67 | 6.5672 (6.1625) | -9.0363 (7.2117) | -1.6301 (1.1530) | 5 | 7 | 15 |
| Total | 203 | 6.8424 (6.5306) | -9.5166 (9.5316) | -1.5470 (1.1546) | 12 | 22 | 41 |
| **Medium-term** | | | | | | | |
| AE | 69 | 12.0725 (8.9234) | -14.6508 (12.5966) | -1.2721 (0.7608) | 2 | 5 | 8 |
| EM | 29 | 14.3103 (10.7008) | -14.2107 (9.6672) | -1.1762 (0.8510) | 1 | 5 | 6 |
| Total | 98 | 12.7347 (9.4831) | -14.5206 (11.7581) | -1.2427 (0.7853) | 3 | 10 | 14 |

Note: This table reports the means of duration, amplitude, and slope, with standard errors in parentheses. The duration of public debt expansion (contraction) is the number of quarters from a trough (peak) to a peak (trough). The amplitude of public debt expansion (contraction) is the percentage growth of the public debt ratio from a

trough (peak) to a peak (trough). Slope = Amplitude / Duration, i.e., the percentage change in the public debt ratio over a quarter. The number of associations with credit booms (busts), house price booms (busts), and equity price booms (busts) is based on the definitions in Section 3.4.

Table 2. Determinants of the duration of public debt expansions in EMs.

| | M1 | M2 | M3 | M4 | M5 | M6 | M7 | M8 |
|---|---|---|---|---|---|---|---|---|
| | | | | Dependent variable: duration of public debt expansions | | | | |
| Credit bust | | -0.1247 | | | 0.0458 | 0.0286 | 0.0168 | -0.0004 |
| | | (0.4370) | | | (0.3936) | (0.3973) | (0.3821) | (0.3851) |
| House price bust | | | 0.6155** | | 0.4797** | 0.4802** | 0.4852** | 0.4850** |
| | | | (0.2470) | | (0.2376) | (0.2372) | (0.2319) | (0.2314) |
| Equity price bust | | | | 0.6267*** | 0.5593*** | 0.5466*** | 0.4508** | 0.4362** |
| | | | | (0.1933) | (0.1937) | (0.1978) | (0.1961) | (0.2011) |
| Credit growth rate | | | | | | 0.0241 | | 0.0239 |
| | | | | | | (0.0760) | | (0.0712) |
| House price growth rate | | | | | | | -0.0731** | -0.0723** |
| | | | | | | | (0.0348) | (0.0343) |
| Constant | 2.2483*** | 2.2567*** | 2.1063*** | 2.0414*** | 1.9559*** | 1.9738*** | 2.0516*** | 2.0696*** |
| | (0.1202) | (0.1230) | (0.1283) | (0.1141) | (0.1218) | (0.1347) | (0.1287) | (0.1397) |
| Weibull shape parameter ($\ln p$) | 0.2220** | 0.2213** | 0.2665** | 0.2621*** | 0.2780*** | 0.2786*** | 0.3097*** | 0.3111*** |
| | (0.1085) | (0.1083) | (0.1059) | (0.0907) | (0.0877) | (0.0875) | (0.0900) | (0.0900) |
| Frailty parameter ($\ln \theta$) | -2.2822 | -2.3351 | -2.1190 | -16.3315 | -15.9006 | -16.1096 | -16.4458 | -15.0146 |
| | (1.4864) | (1.5402) | (1.3014) | (1130.5186) | (1491.3535) | (1694.3427) | (1843.8579) | (905.8436) |
| Log likelihood | -96.0608 | -96.0216 | -92.6078 | -91.5446 | -89.2291 | -89.1790 | -87.1978 | -87.1421 |
| LR Chi-squared | | 0.0783 | 6.9059*** | 9.0324*** | 13.6634*** | 13.7636*** | 17.7260*** | 17.8374*** |
| Number of observations | 72 | 72 | 72 | 72 | 72 | 72 | 72 | 72 |

Note: This table reports the impact of short-term financial busts on the duration of public debt expansions in EMs. All regressions control the shared frailty. Standard errors are in parentheses. $p$ is the Weibull shape parameter, reflecting duration time-dependency. $\theta$ is the variance of $\alpha_i$ in Eq. (4). ** and *** represent

significance at the 5% and 1% level, respectively.

Table 3. Determinants of the duration of public debt cycles.

**Panel A: Determinants of the duration of public debt expansions**

| | Short-term | | | | Medium-term | | | |
|---|---|---|---|---|---|---|---|---|
| | AE | | EM | | AE | | EM | |
| | M5 | M8 | M5 | M8 | M5 | M8 | M5 | M8 |
| Credit bust | 0.2584 | -0.1137 | 0.0458 | -0.0004 | 0.3198 | 0.1097 | 0.7540** | 0.6394* |
| | (0.2585) | (0.2546) | (0.3936) | (0.3851) | (0.2926) | (0.2731) | (0.3388) | (0.3721) |
| House price bust | 0.1978 | 0.1683 | 0.4797** | 0.4850** | 0.2936 | 0.1802 | 0.4513* | 0.4720* |
| | (0.1647) | (0.1535) | (0.2376) | (0.2314) | (0.1918) | (0.1776) | (0.2676) | (0.2668) |
| Equity price bust | 0.1487 | 0.2201* | 0.5593*** | 0.4362** | 0.1357 | 0.1357 | 0.0584 | 0.0311 |
| | (0.1253) | (0.1170) | (0.1937) | (0.2011) | (0.1543) | (0.1423) | (0.2514) | (0.2553) |
| Credit growth rate | | 0.0795* | | 0.0239 | | 0.2503*** | | 0.0495 |
| | | (0.0456) | | (0.0712) | | (0.0612) | | (0.0876) |
| House price growth rate | | -0.1075*** | | -0.0723** | | -0.0217 | | -0.0329 |
| | | (0.0277) | | (0.0343) | | (0.0232) | | (0.0484) |
| Constant | 2.0103*** | 2.0362*** | 1.9559*** | 2.0696*** | 2.5649*** | 2.6729*** | 2.6541*** | 2.7360*** |
| | (0.1010) | (0.1090) | (0.1218) | (0.1397) | (0.1120) | (0.1060) | (0.1861) | (0.2032) |
| Weibull shape parameter ($\ln p$) | 0.4270*** | 0.5149*** | 0.2780*** | 0.3111*** | 0.5529*** | 0.6239*** | 0.6344*** | 0.6206*** |
| | (0.0717) | (0.0734) | (0.0877) | (0.0900) | (0.1131) | (0.1069) | (0.1763) | (0.1764) |
| Frailty parameter ($\ln \theta$) | -1.3647** | -0.7393 | -15.9006 | -15.0146 | -1.4071 | -1.7981 | -1.3522 | -1.8501 |
| | (0.6349) | (0.5647) | (1491.3535) | (905.8436) | (0.9078) | (1.1842) | (1.7058) | (2.3645) |
| Log likelihood | -179.1457 | -171.9588 | -89.2291 | -87.1421 | -88.9945 | -82.0392 | -33.2467 | -32.8861 |
| LR Chi-squared | 3.6417 | 18.0154*** | 13.6634*** | 17.8374*** | 4.1928 | 18.1033*** | 8.5001** | 9.2215* |
| Number of observations | 153 | 153 | 71 | 71 | 84 | 84 | 33 | 33 |

**Panel B: Determinants of the duration of public debt contractions**

| Short-term | Medium-term |
|---|---|

|  | AE | | EM | | AE | | EM | |
|---|---|---|---|---|---|---|---|---|
|  | M5 | M8 | M5 | M8 | M5 | M8 | M5 | M8 |
| Credit boom | 0.4602 | 0.4786 | -0.8328** | -0.9013*** | -0.0048 | 0.0005 | -1.4061*** | -1.4528*** |
|  | (0.2900) | (0.2942) | (0.3407) | (0.3321) | (0.4163) | (0.4175) | (0.4670) | (0.4683) |
| House price boom | 0.0025 | 0.1588 | -0.3642 | -0.4260* | 0.4200* | 0.4284* | 0.6424*** | 0.5828** |
|  | (0.2109) | (0.2248) | (0.2330) | (0.2340) | (0.2350) | (0.2423) | (0.2445) | (0.2436) |
| Equity price boom | 0.5844*** | 0.5988*** | 0.4483* | 0.5185** | 0.0835 | 0.0763 | -0.1675 | 0.0262 |
|  | (0.1610) | (0.1690) | (0.2427) | (0.2396) | (0.1870) | (0.1937) | (0.2410) | (0.2833) |
| Credit growth rate |  | 0.0034 |  | -0.0501** |  | 0.0126 |  | -0.0739 |
|  |  | (0.0640) |  | (0.0226) |  | (0.0941) |  | (0.0477) |
| House price growth rate |  | 0.0953** |  | 0.0003 |  |  |  |  |
|  |  | (0.0482) |  | (0.0383) |  |  |  |  |
| Constant | 1.8514*** | 1.7560*** | 1.9993*** | 2.0828*** | 2.4069*** | 2.3938*** | 2.1315*** | 2.3671*** |
|  | (0.0816) | (0.1131) | (0.1131) | (0.1297) | (0.1207) | (0.1552) | (0.4078) | (0.3627) |
| Weibull shape parameter ($\ln p$) | 0.3304*** | 0.3417*** | 0.3218*** | 0.3586*** | 0.6526*** | 0.6557*** | 0.9387*** | 0.9413*** |
|  | (0.0765) | (0.0771) | (0.0873) | (0.0886) | (0.1154) | (0.1167) | (0.1479) | (0.1674) |
| Frailty parameter ($\ln \theta$) | -3.8717 | -3.9415 | -15.4746 | -15.2358 | -0.6915 | -0.6549 | 1.9298 | 1.3967 |
|  | (4.2039) | (4.1771) | (1670.0177) | (1339.2873) | (0.8088) | (0.8386) | (1.5231) | (1.5398) |
| Log likelihood | -162.3003 | -160.3382 | -80.4265 | -78.7520 | -70.7955 | -70.7866 | -28.4988 | -27.4478 |
| LR Chi-squared | 14.2457*** | 18.1699*** | 9.3879** | 12.7371** | 4.8186 | 4.8364 | 8.2476** | 10.3497** |
| Number of observations | 136 | 136 | 67 | 67 | 69 | 69 | 29 | 29 |

Note: This table reports the impact of financial cycles on the duration of public debt cycles. All regressions control the shared frailty. Standard errors are in parentheses. $p$ is the Weibull shape parameter, reflecting duration time-dependency. $\theta$ is the variance of $\alpha_i$ in Eq. (4). "Short-term" ("Medium-term") columns represent the regressions after applying short-term (medium-term) censoring rules. The house price growth rate is dropped from Panel B because it causes significant multi-collinearity in the medium-term contraction regressions. *, ** and *** represent significance at the 10%, 5% and 1% level, respectively.

Table 4. Determinants of the amplitude of public debt cycles.

**Panel A: Determinants of the amplitude of public debt expansions**

| | Short-term | | | | Medium-term | | | |
| --- | --- | --- | --- | --- | --- | --- | --- | --- |
| | AE | | EM | | AE | | EM | |
| | M5 | M8 | M5 | M8 | M5 | M8 | M5 | M8 |
| Credit bust | 19.6782 | 1.0058 | -3.5882 | -7.5033 | 40.8319 | 45.4913 | 67.5407* | 58.6664* |
| | (15.7653) | (16.0408) | (22.3197) | (21.2188) | (31.7946) | (30.6047) | (32.8067) | (32.0013) |
| House price bust | -15.6615 | -19.4834* | 29.4702** | 27.2556** | -13.4273 | -18.0918 | 27.5894 | 39.5535 |
| | (10.7793) | (10.4053) | (13.3920) | (12.7970) | (21.9975) | (21.6446) | (25.2932) | (25.0747) |
| Equity price bust | 5.5140 | 12.5913* | 17.7395 | 12.4718 | -2.1803 | -5.2237 | 9.1530 | 24.8124 |
| | (7.5654) | (7.5393) | (11.6770) | (11.3449) | (17.3230) | (16.7175) | (32.8067) | (32.5093) |
| Credit growth rate | | 7.2604** | | 0.9942 | | 20.7409** | | 2.0393 |
| | | (2.8162) | | (3.8807) | | (8.4212) | | (12.7064) |
| House price growth rate | | -5.0474** | | -7.3342*** | | -5.3979 | | -11.5455* |
| | | (1.9862) | | (2.4297) | | (4.5627) | | (6.4465) |
| Constant | 28.6446*** | 32.2401*** | 41.5568*** | 51.3561*** | 51.5488*** | 61.6700*** | 90.1043*** | 94.3740*** |
| | (4.5902) | (4.5498) | (6.4849) | (7.3810) | (9.3752) | (10.2199) | (12.9125) | (14.0130) |
| Log likelihood | -769.4425 | -761.9499 | -352.5580 | -346.4171 | -445.9797 | -440.9675 | -158.9844 | -154.9046 |
| Number of observations | 153 | 153 | 71 | 71 | 84 | 84 | 33 | 33 |

**Panel B: Determinants of the amplitude of public debt contractions**

| | Short-term | | | | Medium-term | | | |
| --- | --- | --- | --- | --- | --- | --- | --- | --- |
| | AE | | EM | | AE | | EM | |
| | M5 | M8 | M5 | M8 | M5 | M8 | M5 | M8 |
| Credit boom | -1.6237 | -2.4221 | 6.2385* | 6.4203* | 10.1944 | 11.6417 | 15.7882* | 15.2818* |
| | (4.6734) | (4.7514) | (3.8064) | (3.8960) | (10.3316) | (10.9063) | (7.9599) | (8.3323) |

| | (1) | (2) | (3) Short-term | (4) Short-term | (5) Medium-term | (6) Medium-term | (7) | (8) |
|---|---|---|---|---|---|---|---|---|
| House price boom | 1.3308 | 0.6476 | -2.4672 | -2.6559 | -6.5914 | -5.7685 | -12.2354** | -12.5438* |
| | (3.2142) | (3.5132) | (2.9458) | (3.0294) | (6.1469) | (6.9935) | (4.0898) | (5.9235) |
| Equity price boom | -5.3501** | -5.3379** | -2.4094 | -2.5377 | -3.6714 | -3.5650 | 6.6463 | 8.3441 |
| | (2.3120) | (2.3225) | (2.6567) | (2.7375) | (4.9033) | (5.1489) | (4.2120) | (4.7252) |
| Credit growth rate | | -0.7610 | | 0.1141 | | -2.0406 | | -1.2860 |
| | | (0.6707) | | (0.3289) | | (2.1515) | | (1.1966) |
| House price growth rate | | -0.7934 | | -0.2296 | | 1.6732 | | -0.0629 |
| | | (0.7334) | | (0.4599) | | (1.5268) | | (0.9077) |
| Constant | -8.5872*** | -7.2409*** | -8.6644*** | -8.6168*** | -13.0865*** | -12.8400*** | -13.6352*** | -11.6577*** |
| | (1.1126) | (1.4204) | (1.1449) | (1.3418) | (1.8362) | (2.9845) | (1.6874) | (3.1447) |
| Log likelihood | -495.0341 | -493.5656 | -212.4825 | -212.2310 | -250.3651 | -248.4299 | -78.9373 | -76.9625 |
| Number of observations | 136 | 136 | 67 | 67 | 69 | 69 | 29 | 29 |

Note: This table reports the impact of financial cycles on the amplitude of public debt cycles. All regressions control the country's fixed effect. Standard errors are in parentheses. "Short-term" ("Medium-term") columns represent the regressions after applying short-term (medium-term) censoring rules. *, ** and *** represent significance at the 10%, 5% and 1% level, respectively.

Table 5. Determinants of the duration of public debt cycles: Robustness (PCA test).

**Panel A: Robustness test for the determinants of public debt expansions**

|  | Short-term | | Medium-term | |
|---|---|---|---|---|
|  | AE | EM | AE | EM |
|  | (I) | (II) | (III) | (IV) |
| Credit bust | 0.0121 | 0.1182 | 0.0752 | 0.8728** |
|  | (0.2710) | (0.3865) | (0.2575) | (0.3520) |
| House price bust | 0.1988 | 0.4004* | 0.1532 | 0.6466** |
|  | (0.1451) | (0.2343) | (0.1716) | (0.2821) |
| Equity price bust | 0.1866* | 0.5322** | 0.0760 | 0.0363 |
|  | (0.1105) | (0.2075) | (0.1366) | (0.2583) |
| Credit growth rate | 0.0580 | -0.0015 | 0.1952*** | 0.0980 |
|  | (0.0398) | (0.0788) | (0.0667) | (0.1267) |
| House price growth rate | -0.0983*** | -0.0611* | -0.0440 | -0.0229 |
|  | (0.0346) | (0.0345) | (0.0311) | (0.0491) |
| PC1 | 0.0537 | -0.0285 | 0.1074 | -0.1017 |
|  | (0.0528) | (0.0881) | (0.0702) | (0.0900) |
| PC2 | -0.1409** | 0.1366 | 0.0976* | 0.1681* |
|  | (0.0683) | (0.0871) | (0.0521) | (0.1021) |
| PC3 | 0.2010*** | 0.0265 | -0.1483** | 0.1762 |
|  | (0.0536) | (0.1007) | (0.0726) | (0.1624) |
| Constant | 2.0006*** | 2.0187*** | 2.6613*** | 2.6033*** |
|  | (0.1016) | (0.1424) | (0.1059) | (0.1940) |
| Weibull shape parameter (ln $p$) | 0.5932*** | 0.3366*** | 0.6900*** | 0.7684*** |
|  | (0.0697) | (0.0918) | (0.1014) | (0.1802) |
| Frailty parameter (ln $\theta$) | -0.8070 | -14.7402 | -1.4993* | -0.7262 |
|  | (0.6045) | (834.6739) | (0.9062) | (1.3644) |
| Log likelihood | -160.2313 | -85.9557 | -76.3952 | -30.3475 |
| LR Chi-squared | 41.4706*** | 20.2102*** | 29.3914*** | 14.2986* |
| Number of observations | 153 | 71 | 84 | 32 |

**Panel B: Robustness test for the determinants of public debt contractions**

|  | Short-term | | Medium-term | |
|---|---|---|---|---|
|  | AE | EM | AE | EM |
|  | (I) | (II) | (III) | (IV) |
| Credit boom | 0.5477* | -0.9327*** | 0.0401 | -1.0897** |
|  | (0.2810) | (0.3460) | (0.4028) | (0.5497) |
| House price boom | 0.1453 | -0.3939* | 0.5675** | 0.5883** |
|  | (0.2133) | (0.2448) | (0.2529) | (0.2632) |
| Equity price boom | 0.5830*** | 0.5079* | 0.1003 | 0.2725 |
|  | (0.1463) | (0.2830) | (0.1899) | (0.3540) |
| Credit growth rate | -0.0211 | -0.0663 | 0.0072 | 0.0317 |
| | (0.0583) | (0.0466) | (0.0936) | (0.0995) |
| House price growth rate | 0.1123** | 0.0009 | | |
| | (0.0467) | (0.0390) | | |
| PC1 | -0.0176 | 0.0443 | -0.0828 | 0.0317 |
| | (0.0534) | (0.1236) | (0.0733) | (0.1015) |
| PC2 | -0.1192* | -0.0148 | 0.1463** | 0.2309 |
| | (0.0696) | (0.0828) | (0.0735) | (0.1517) |
| PC3 | 0.0781 | -0.0472 | 0.1603 | 0.0466 |
| | (0.0792) | (0.0955) | (0.1117) | (0.1396) |
| Constant | 1.7697*** | 2.1013*** | 2.3653*** | 2.3454*** |
| | (0.1042) | (0.1369) | (0.1576) | (0.2638) |
| Weibull shape parameter (ln $p$) | 0.3632*** | 0.3561*** | 0.6904*** | 0.8736*** |
| | (0.0629) | (0.0902) | (0.1210) | (0.1798) |
| Frailty parameter (ln $\theta$) | -16.0491 | -16.3856 | -0.8235 | 0.3056 |
| | (1130.0717) | (1834.7862) | (0.9462) | (1.2861) |
| Log likelihood | -157.8868 | -77.9206 | -67.3335 | -25.5797 |
| LR Chi-squared | 23.0727*** | 12.4902 | 11.7424* | 10.3920 |
| Number of observations | 136 | 66 | 69 | 28 |

Note: This table reports the robustness tests for the impact of financial cycles on the duration of public debt cycles. All regressions control the shared frailty. Standard errors are in parentheses. $p$ is the Weibull shape parameter, reflecting duration time-dependency. $\theta$ is the variance of $\alpha_i$ in Eq. (4). "Short-term" ("Medium-term") columns represent the regressions after applying short-term (medium-term) censoring rules. House price growth is not included in medium-term contraction regressions, i.e., columns III and IV in Panel B, because it causes severe multi-collinearity. PC1, PC2 and PC3 are the first, second and third primary components of the six macro variables. *, ** and *** represent significance at the 10%, 5% and 1% level, respectively.

# Appendix A

Appendix A summarizes the complete sample coverage and data sources of this paper. Table A.1 shows the starting date of data for each country. All data end in 2022Q4.

Table A.1. Starting date of data.

| AE | Starting date | EM | Starting date |
| --- | --- | --- | --- |
| Australia | 1988Q2 | Argentina | 2006Q4 |
| Austria | 1999Q4 | Brazil | 2001Q3 |
| Belgium | 2003Q1 | Chile | 2003Q2 |
| Canada | 1990Q1 | China | 1998Q1 |
| Czech Republic | 2007Q1 | Colombia | 1996Q4 |
| Denmark | 2005Q1 | Hungary | 1995Q2 |
| Finland | 1998Q4 | India | 2002Q1 |
| France | 1998Q4 | Indonesia | 2001Q4 |
| Germany | 1998Q4 | Malaysia | 1999Q3 |
| Greece | 2002Q1 | Mexico | 2006Q2 |
| Ireland | 1999Q4 | Philippines | 2007Q4 |
| Israel | 2002Q1 | Poland | 2005Q3 |
| Italy | 1999Q1 | Romania | 2007Q3 |
| Japan | 1997Q4 | Russia | 2003Q1 |
| Luxembourg | 1998Q4 | Saudi Arabia | 2013Q4 |
| Netherlands | 2003Q2 | Thailand | 1997Q1 |
| New Zealand | 1989Q4 | The Republic of South Africa | 1988Q1 |
| Norway | 1995Q4 | Turkey | 2010Q1 |
| Portugal | 1998Q4 | | |
| Singapore | 1993Q3 | | |
| South Korea | 1990Q4 | | |
| Spain | 1995Q1 | | |
| Sweden | 1999Q1 | | |
| Switzerland | 1995Q4 | | |
| the U.K. | 1990Q1 | | |
| the U.S. | 1990Q1 | | |

Table A.2 shows the data sources.

Table A.2. Data sources.

| Variable | Source | Description |
| --- | --- | --- |
| Public debt to GDP ratio | BIS statistics and CEIC database | Data of most AEs are from BIS. |
| Private debt to GDP ratio | BIS statistics and CEIC database | When BIS data are not available |
| | | (usually some EMs), we obtain them from CEIC database. |
| House price | OECD, BIS and CEIC database | Data of most AEs are real house price index from OECD. When OECD data are not available (usually some EMs), we obtain them from CEIC database. |
| Equity price | OECD, CEIC database and national sources | Data of most AEs are share price index from OECD. When OECD data are not available (usually some EMs), we obtain them from CEIC database. |
| GDP (growth rate) Monetary supply (M2 or M3 growth rate) Inflation (CPI growth rate, last quarter) REER index Account Balance Crude oil price (NYMEX crude oil price growth rate) | OECD, IMF (IFS datasets), CEIC database | When OECD and IMF data are not available (usually some EMs), we obtain them from CEIC database. |

Note: BIS statistics: https://www.bis.org/statistics. OECD data: https://data.oecd.org. IMF data: https://data.imf.org CEIC database: https://www.ceicdata.com.cn.

We also include six macroeconomic variables for robustness tests, i.e., real GDP growth rate, money supply growth rate, inflation rate, account balance, REER growth rate and crude oil price growth rate. The real GDP growth rate reflects the overall economic condition, measured by the quarter-to-quarter change in real GDP. The money supply growth rate is measured by the change in the broad money supply[1]. The inflation rate is the quarter-to-quarter change in the Consumer Price Index (CPI). Account balance is the percentage of the current account balance-to-GDP ratio. REER growth rate represents the change in the Real Effective Exchange Rate (REER) index based on CPI. The crude oil price growth rate is the change in the NYMEX crude oil price. All these macroeconomic variables are in percentage terms and they directly or indirectly impact fiscal balances and public debt levels. Table A.3 summarizes the definitions and calculations of all variables that are eventually taken into models after cycle identifications.

---

[1] For AEs and most EMs, we use M3 to represent broad money supply, whereas for some less developed EMs where M3 data is unavailable, we use M2 instead.

**Table A.3. Variable introduction.**

| Variable | Type | Definition |
|---|---|---|
| *Dummy variables* | | |
| Credit bust (boom) | Dummy variable: take one when public debt expansions (contractions) are associated with credit busts (booms), otherwise zero. | See *Section 3.4* |
| House bust (boom) | Dummy variable: take one when public debt expansions (contractions) Are associated with house price busts (booms), otherwise zero. | |
| Equity bust (boom) | Dummy variable: take one when public debt expansions (contractions) Are associated with equity price busts (booms), otherwise zero. | |
| *Financial growth variables* | | |
| Credit growth | Private sector credit to GDP ratio, percentage change. | Two-quarters average growth before each public debt peak (trough). |
| House price growth | Real house price index, percentage change. | Two-quarters average growth before each public debt peak (trough). |
| *Macro variables* | | |
| Real GDP growth | Real GDP, percentage change. | One quarter's average growth before public debt peak (trough). |
| Monetary supply growth | Aggregate money supply, percentage change. | Two-quarters average growth before each public debt peak (trough). |
| Inflation | CPI, percentage change (last period). | Two-quarters average growth after each public debt peak (trough). |
| REER growth | Real effective exchange rate Index based on CPI, percentage change. | Two-quarters average growth before each public debt peak (trough). |
| Current account balance | Current account balance to GDP ratio, in percentage. | Two-quarters average before each public debt peak (trough). |
| Crude oil price growth | NYMEX crude oil price, percentage change. | Three-quarters average growth before each public debt peak and trough. |

Note: The average growth is the arithmetic mean of percentage change in variables. GDP, monetary supply and inflation represent the main domestic economic indicators. The current account balance reflects the government's

fiscal status. REER index and crude oil price show the external impact of the international situation and energy costs.

# Appendix B

Appendix B summarizes the results for the impact of financial cycles on the duration and amplitude of public debt cycles. The results are divided into eight groups (panels), i.e., how financial busts (booms) impact public debt expansions (contractions) in short-term and medium-term cycles, and in advanced economies (AEs) and emerging economies (EMs).

Table B.1 presents the results based on the Weibull AFT survival model with shared frailty for the impact of financial cycles on the duration of public debt cycles. M1 is the benchmark model. M2 to M4 examine the impact of financial cycles respectively. M5 includes the three dummy variables, and M6-M8 further consider credit growth rate and house price growth rate.

Table B.1. Determinants of the duration of public debt cycles.

**Panel A: Determinants of the duration of short-term public debt expansions in AEs**

| | M1 | M2 | M3 | M4 | M5 | M6 | M7 | M8 |
|---|---|---|---|---|---|---|---|---|
| Credit bust | | 0.1825 | | | 0.2584 | 0.1403 | 0.0295 | -0.1137 |
| | | (0.2534) | | | (0.2585) | (0.2669) | (0.2483) | (0.2546) |
| House price bust | | | 0.2029 | | 0.1978 | 0.1721 | 0.2024 | 0.1683 |
| | | | (0.1648) | | (0.1647) | (0.1635) | (0.1557) | (0.1535) |
| Equity price bust | | | | 0.1308 | 0.1487 | 0.1797 | 0.1839 | 0.2201* |
| | | | | (0.1228) | (0.1253) | (0.1260) | (0.1167) | (0.1170) |
| Credit growth rate | | | | | | 0.0749 | | 0.0795* |
| | | | | | | (0.0492) | | (0.0456) |
| House price growth rate | | | | | | | -0.1039*** | -0.1075*** |
| | | | | | | | (0.0280) | (0.0277) |
| Constant | 2.1041*** | 2.0954*** | 2.0732*** | 2.0583*** | 2.0103*** | 2.0187*** | 2.0283*** | 2.0362*** |
| | (0.0931) | (0.0932) | (0.0944) | (0.1000) | (0.1010) | (0.0992) | (0.1101) | (0.1090) |
| Weibull shape parameter (ln $p$) | 0.4282*** | 0.4303*** | 0.4266*** | 0.4268*** | 0.4270*** | 0.4340*** | 0.5010*** | 0.5149*** |
| | (0.0699) | (0.0703) | (0.0715) | (0.0697) | (0.0717) | (0.0704) | (0.0752) | (0.0734) |
| Frailty parameter (ln $\theta$) | -1.1489** | -1.1604** | -1.2245** | -1.2345** | -1.3647** | -1.4377** | -0.7476 | -0.7393 |
| | (0.5810) | (0.5845) | (0.6060) | (0.5927) | (0.6349) | (0.6536) | (0.5697) | (0.5647) |
| Log likelihood | -180.9666 | -180.6947 | -180.1687 | -180.3935 | -179.1457 | -177.9076 | -173.6062 | -171.9588 |
| LR Chi-squared | | 0.5437 | 1.5958 | 1.1461 | 3.6417 | 6.1180 | 14.7207*** | 18.0154*** |
| Number of observations | 153 | 153 | 153 | 153 | 153 | 153 | 153 | 153 |

**Panel B: Determinants of the duration of short-term public debt expansions in EMs**

| | M1 | M2 | M3 | M4 | M5 | M6 | M7 | M8 |
|---|---|---|---|---|---|---|---|---|
| Credit bust | | -0.1247 | | | 0.0458 | 0.0286 | 0.0168 | -0.0004 |
| | | (0.4370) | | | (0.3936) | (0.3973) | (0.3821) | (0.3851) |
| House price bust | | | 0.6155** | | 0.4797** | 0.4802** | 0.4852** | 0.4850** |
| | | | (0.2470) | | (0.2376) | (0.2372) | (0.2319) | (0.2314) |

| | | | | 0.6267*** | 0.5593*** | 0.5466*** | 0.4508** | 0.4362** |
|---|---|---|---|---|---|---|---|---|
| Equity price bust | | | | (0.1933) | (0.1937) | (0.1978) | (0.1961) | (0.2011) |
| Credit growth rate | | | | | | 0.0241 | | 0.0239 |
| | | | | | | (0.0760) | | (0.0712) |
| House price growth rate | | | | | | | -0.0731** | -0.0723** |
| | | | | | | | (0.0348) | (0.0343) |
| Constant | 2.2483*** | 2.2567*** | 2.1063*** | 2.0414*** | 1.9559*** | 1.9738*** | 2.0516*** | 2.0696*** |
| | (0.1202) | (0.1230) | (0.1283) | (0.1141) | (0.1218) | (0.1347) | (0.1287) | (0.1397) |
| Weibull shape parameter (ln $p$) | 0.2220** | 0.2213** | 0.2665** | 0.2621*** | 0.2780*** | 0.2786*** | 0.3097*** | 0.3111*** |
| | (0.1085) | (0.1083) | (0.1059) | (0.0907) | (0.0877) | (0.0875) | (0.0900) | (0.0900) |
| Frailty parameter (ln $\theta$) | -2.2822 | -2.3351 | -2.1190 | -16.3315 | -15.9006 | -16.1096 | -16.4458 | -15.0146 |
| | (1.4864) | (1.5402) | (1.3014) | (1130.5186) | (1491.3535) | (1694.3427) | (1843.8579) | (905.8436) |
| Log likelihood | -96.0608 | -96.0216 | -92.6078 | -91.5446 | -89.2291 | -89.1790 | -87.1978 | -87.1421 |
| LR Chi-squared | | 0.0783 | 6.9059*** | 9.0324*** | 13.6634*** | 13.7636*** | 17.7260*** | 17.8374*** |
| Number of observations | 72 | 72 | 72 | 72 | 72 | 72 | 72 | 72 |

**Panel C: Determinants of the duration of medium-term public debt expansions in AEs**

| | M1 | M2 | M3 | M4 | M5 | M6 | M7 | M8 |
|---|---|---|---|---|---|---|---|---|
| Credit bust | | 0.2358 | | | 0.3198 | 0.0904 | 0.3379 | 0.1097 |
| | | (0.2968) | | | (0.2926) | (0.2721) | (0.2956) | (0.2731) |
| House price bust | | | 0.2882 | | 0.2936 | 0.1965 | 0.2866 | 0.1802 |
| | | | (0.1928) | | (0.1918) | (0.1773) | (0.1927) | (0.1776) |
| Equity price bust | | | | 0.1412 | 0.1357 | 0.1419 | 0.1318 | 0.1357 |
| | | | | (0.1559) | (0.1543) | (0.1425) | (0.1546) | (0.1423) |
| Credit growth rate | | | | | | 0.2407*** | | 0.2503*** |
| | | | | | | (0.0598) | | (0.0612) |
| House price growth rate | | | | | | | -0.0130 | -0.0217 |
| | | | | | | | (0.0265) | (0.0232) |
| Constant | 2.6775*** | 2.6579*** | 2.6341*** | 2.6358*** | 2.5649*** | 2.6567*** | 2.5722*** | 2.6729*** |
|  | (0.1021) | (0.1049) | (0.1005) | (0.1091) | (0.1120) | (0.1046) | (0.1130) | (0.1060) |
| Weibull shape parameter (ln $p$) | 0.5434*** | 0.5534*** | 0.5379*** | 0.5406*** | 0.5529*** | 0.6196*** | 0.5535*** | 0.6239*** |
|  | (0.1045) | (0.1047) | (0.1111) | (0.1053) | (0.1131) | (0.1077) | (0.1130) | (0.1069) |
| Frailty parameter (ln $\theta$) | -1.1882 | -1.1233 | -1.4365 | -1.2864 | -1.4071 | -1.8180 | -1.3931 | -1.7981 |
|  | (0.8738) | (0.8464) | (0.9584) | (0.8855) | (0.9078) | (1.1668) | (0.9084) | (1.1842) |
| Log likelihood | -91.0909 | -90.7554 | -89.9197 | -90.6759 | -88.9945 | -82.4593 | -88.8762 | -82.0392 |
| LR Chi-squared |  | 0.6709 | 2.3424 | 0.8300 | 4.1928 | 17.2631*** | 4.4294 | 18.1033*** |
| Number of observations | 84 | 84 | 84 | 84 | 84 | 84 | 84 | 84 |

**Panel D: Determinants of the duration of medium-term public debt expansions in EMs**

|  | M1 | M2 | M3 | M4 | M5 | M6 | M7 | M8 |
|---|---|---|---|---|---|---|---|---|
| Credit bust |  | 0.6839** |  |  | 0.7540** | 0.7510** | 0.6398* | 0.6394* |
|  |  | (0.3432) |  |  | (0.3388) | (0.3332) | (0.3796) | (0.3721) |
| House price bust |  |  | 0.4101 |  | 0.4513* | 0.4398* | 0.4784* | 0.4720* |
|  |  |  | (0.2806) |  | (0.2676) | (0.2638) | (0.2702) | (0.2668) |
| Equity price bust |  |  |  | 0.1337 | 0.0584 | 0.0177 | 0.0744 | 0.0311 |
|  |  |  |  | (0.2803) | (0.2514) | (0.2595) | (0.2468) | (0.2553) |
| Credit growth rate |  |  |  |  |  | 0.0484 |  | 0.0495 |
|  |  |  |  |  |  | (0.0895) |  | (0.0876) |
| House price growth rate |  |  |  |  |  |  | -0.0333 | -0.0329 |
|  |  |  |  |  |  |  | (0.0502) | (0.0484) |
| Constant | 2.9496*** | 2.8260*** | 2.8341*** | 2.9014*** | 2.6541*** | 2.6956*** | 2.6968*** | 2.7360*** |
|  | (0.1330) | (0.1594) | (0.1506) | (0.1674) | (0.1861) | (0.2001) | (0.1937) | (0.2032) |
| Weibull shape parameter (ln $p$) | 0.4440*** | 0.5493*** | 0.4848*** | 0.4620*** | 0.6344*** | 0.6328*** | 0.6185*** | 0.6206*** |
|  | (0.1675) | (0.1994) | (0.1631) | (0.1693) | (0.1763) | (0.1778) | (0.1753) | (0.1764) |
| Frailty parameter (ln $\theta$) | -3.2056 | -2.1955 | -2.4652 | -2.5351 | -1.3522 | -1.5194 | -1.6926 | -1.8501 |
|  | (5.2983) | (3.6859) | (2.6566) | (3.0856) | (1.7058) | (1.9476) | (2.0589) | (2.3645) |
| Log likelihood | -37.4968 | -34.9659 | -36.3076 | -37.3793 | -33.2467 | -33.1070 | -33.0374 | -32.8861 |
| LR Chi-squared |  | 5.0619** | 2.3783 | 0.2351 | 8.5001** | 8.7797* | 8.9187* | 9.2215* |
| Number of observations | 33 | 33 | 33 | 33 | 33 | 33 | 33 | 33 |

**Panel E: Determinants of the duration of short-term public debt contractions in AEs**

| | M1 | M2 | M3 | M4 | M5 | M6 | M7 | M8 |
|---|---|---|---|---|---|---|---|---|
| Credit boom | | 0.3356 | | | 0.4602 | 0.4500 | 0.4752 | 0.4786 |
| | | (0.3208) | | | (0.2900) | (0.2964) | (0.2871) | (0.2942) |
| House price boom | | | 0.0915 | | 0.0025 | -0.0009 | 0.1572 | 0.1588 |
| | | | (0.2254) | | (0.2109) | (0.2118) | (0.2230) | (0.2248) |
| Equity price boom | | | | 0.5466*** | 0.5844*** | 0.5860*** | 0.5994*** | 0.5988*** |
| | | | | (0.1622) | (0.1610) | (0.1610) | (0.1686) | (0.1690) |
| Credit growth rate | | | | | | -0.0102 | | 0.0034 |
| | | | | | | (0.0629) | | (0.0640) |
| House price growth rate | | | | | | | 0.0950** | 0.0953** |
| | | | | | | | (0.0478) | (0.0482) |
| Constant | 1.9959*** | 1.9819*** | 1.9857*** | 1.8812*** | 1.8514*** | 1.8618*** | 1.7596*** | 1.7560*** |
| | (0.0905) | (0.0899) | (0.0941) | (0.0804) | (0.0816) | (0.1036) | (0.0902) | (0.1131) |
| Weibull shape parameter ($\ln p$) | 0.3136*** | 0.3180*** | 0.3164*** | 0.3181*** | 0.3304*** | 0.3287*** | 0.3412*** | 0.3417*** |
| | (0.0762) | (0.0757) | (0.0761) | (0.0779) | (0.0765) | (0.0776) | (0.0768) | (0.0771) |
| Frailty parameter ($\ln \theta$) | -1.8814** | -1.9270** | -1.8515** | -3.4175 | -3.8717 | -4.0443 | -3.9825 | -3.9415 |
| | (0.8548) | (0.8633) | (0.8368) | (2.8677) | (4.2039) | (5.0763) | (4.2697) | (4.1771) |
| Log likelihood | -169.4232 | -168.8337 | -169.3391 | -163.7574 | -162.3003 | -162.2873 | -160.3396 | -160.3382 |
| LR Chi-squared | | 1.1790 | 0.1681 | 11.3316*** | 14.2457*** | 14.2717*** | 18.1672*** | 18.1699*** |
| Number of observations | 136 | 136 | 136 | 136 | 136 | 136 | 136 | 136 |

**Panel F: Determinants of the duration of short-term public debt contractions in EMs**

| | M1 | M2 | M3 | M4 | M5 | M6 | M7 | M8 |
|---|---|---|---|---|---|---|---|---|
| Credit boom | | -0.9320*** | | | -0.8328** | -0.9016*** | -0.8346** | -0.9013*** |
| | | (0.3504) | | | (0.3407) | (0.3305) | (0.3423) | (0.3321) |
| House price boom | | | -0.2949 | | -0.3642 | -0.4263* | -0.3672 | -0.4260* |
| | | | (0.2515) | | (0.2330) | (0.2274) | (0.2403) | (0.2340) |
| Equity price boom |  |  |  | 0.4564* (0.2589) | 0.4483* (0.2427) | 0.5182** (0.2356) | 0.4462* (0.2460) | 0.5185** (0.2396) |
| Credit growth rate |  |  |  |  |  | -0.0501** (0.0226) |  | -0.0501** (0.0226) |
| House price growth rate |  |  |  |  |  |  | -0.0020 (0.0395) | 0.0003 (0.0383) |
| Constant | 1.9418*** (0.1228) | 1.9913*** (0.1219) | 2.0142*** (0.1126) | 1.8675*** (0.1237) | 1.9993*** (0.1131) | 2.0832*** (0.1175) | 2.0022*** (0.1268) | 2.0828*** (0.1297) |
| Weibull shape parameter ($\ln p$) | 0.2576** (0.1128) | 0.3174*** (0.0999) | 0.2374*** (0.0884) | 0.2778*** (0.1064) | 0.3218*** (0.0873) | 0.3587*** (0.0880) | 0.3224*** (0.0879) | 0.3586*** (0.0886) |
| Frailty parameter ($\ln \theta$) | -2.9561 (2.6070) | -2.4777 (1.6118) | -14.6835 (2117.7618) | -3.4728 (3.7250) | -15.4746 (1670.0177) | -15.1467 (1274.3827) | -15.4677 (1615.8063) | -15.2358 (1339.2873) |
| Log likelihood | -85.1205 | -82.6766 | -84.5988 | -83.4577 | -80.4265 | -78.7520 | -80.4252 | -78.7520 |
| LR LR Chi-squared |  | 4.8879** | 1.0435 | 3.3256* | 9.3879** | 12.7370** | 9.3906** | 12.7371** |
| Number of observations | 67 | 67 | 67 | 67 | 67 | 67 | 67 | 67 |

**Panel G: Determinants of the duration of medium-term public debt contractions in AEs**

|  | M1 | M2 | M3 | M4 | M5 | M6 |
|---|---|---|---|---|---|---|
| Credit boom |  | -0.1980 (0.4275) |  |  | -0.0048 (0.4163) | 0.0005 (0.4175) |
| House price boom |  |  | 0.4583** (0.2144) |  | 0.4200* (0.2350) | 0.4284* (0.2423) |
| Equity price boom |  |  |  | 0.2203 (0.1808) | 0.0835 (0.1870) | 0.0763 (0.1937) |
| Credit growth rate |  |  |  |  |  | 0.0126 (0.0941) |
| House price growth rate |  |  |  |  |  |  |
| Constant | 2.4966*** | 2.4947*** | 2.4190*** | 2.4439*** | 2.4069*** | 2.3938*** |
|  | (0.1153) | (0.1167) | (0.1184) | (0.1206) | (0.1207) | (0.1552) |
| Weibull shape parameter (ln p) | 0.5874*** | 0.5992*** | 0.6573*** | 0.5953*** | 0.6526*** | 0.6557*** |
|  | (0.1170) | (0.1181) | (0.1144) | (0.1157) | (0.1154) | (0.1167) |
| Frailty parameter (ln θ) | -0.8737 | -0.7773 | -0.6364 | -0.9267 | -0.6915 | -0.6549 |
|  | (0.8406) | (0.8278) | (0.7838) | (0.8551) | (0.8088) | (0.8386) |
| Log likelihood | -73.2048 | -73.1096 | -70.8958 | -72.4458 | -70.7955 | -70.7866 |
| LR Chi-squared |  | 0.1903 | 4.6179** | 1.5178 | 4.8186 | 4.8364 |
| Number of observations | 69 | 69 | 69 | 69 | 69 | 69 |

**Panel H: Determinants of the duration of medium-term public debt contractions in EMs**

|  | M1 | M2 | M3 | M4 | M5 | M6 |
|---|---|---|---|---|---|---|
| Credit boom |  | -1.3580** |  |  | -1.4061*** | -1.4528*** |
|  |  | (0.6274) |  |  | (0.4670) | (0.4683) |
| House price boom |  |  | 0.5608** |  | 0.6424*** | 0.5828** |
|  |  |  | (0.2550) |  | (0.2445) | (0.2436) |
| Equity price boom |  |  |  | 0.0876 | -0.1675 | 0.0262 |
|  |  |  |  | (0.2814) | (0.2410) | (0.2833) |
| Credit growth rate |  |  |  |  |  | -0.0739 |
|  |  |  |  |  |  | (0.0477) |
| House price growth rate |  |  |  |  |  |  |
| Constant | 2.7127*** | 2.7200*** | 2.3281*** | 2.6871*** | 2.1315*** | 2.3671*** |
|  | (0.1852) | (0.2178) | (0.3484) | (0.1998) | (0.4078) | (0.3627) |
| Weibull shape parameter (ln p) | 0.4895** | 0.5639** | 0.7429*** | 0.4934** | 0.9387*** | 0.9413*** |
|  | (0.2057) | (0.2199) | (0.2180) | (0.2043) | (0.1479) | (0.1674) |
| Frailty parameter (ln θ) | -1.6307 | -1.2265 | 0.5340 | -1.6069 | 1.9298 | 1.3967 |
|  | (2.3447) | (2.3024) | (1.6665) | (2.2685) | (1.5231) | (1.5398) |
| Log likelihood | -32.6226 | -31.2986 | -30.7213 | -32.5736 | -28.4988 | -27.4478 |
| LR Chi-squared |  | 2.6480* | 3.8025** | 0.0980 | 8.2476** | 10.3497** |

| Number of observations | 29 | 29 | 29 | 29 | 29 | 29 |

Note: This table reports the impact of financial cycles on the duration of public debt cycles. All regressions control the shared frailty. Standard errors are in parentheses. $p$ is the Weibull shape parameter, reflecting duration time-dependency. $\theta$ is the variance of $\alpha_i$ in Eq. (4). The house price growth rate is dropped from medium-term contraction regressions because it causes significant multi-collinearity. *, ** and *** represent significance at the 10%, 5% and 1% level, respectively.

Table B.2 presents the results of the impact of financial cycles on the amplitude of public debt cycles based on panel regressions with fixed effects. M1 is the benchmark model. M2 to M4 examine the impact of financial cycles respectively. M5 includes the three dummy variables, and M6-M8 further consider credit growth rate and house price growth rate.

Table B.2. Determinants of the amplitude of public debt cycles.

**Panel A: Determinants of the amplitude of short-term public debt expansions in AEs**

|  | M1 | M2 | M3 | M4 | M5 | M6 | M7 | M8 |
|---|---|---|---|---|---|---|---|---|
| Credit bust |  | 20.5635 |  |  | 19.6782 | 9.3795 | 12.1215 | 1.0058 |
|  |  | (15.6445) |  |  | (15.7653) | (16.0432) | (15.8011) | (16.0408) |
| House price bust |  |  | -15.8392 |  | -15.6615 | -17.8021* | -17.1757 | -19.4834* |
|  |  |  | (10.6399) |  | (10.7793) | (10.6121) | (10.6020) | (10.4053) |
| Equity price bust |  |  |  | 2.9714 | 5.5140 | 8.3779 | 9.4164 | 12.5913* |
|  |  |  |  | (7.5156) | (7.5654) | (7.5161) | (7.6068) | (7.5393) |
| Credit growth rate |  |  |  |  |  | 6.9466** |  | 7.2604** |
|  |  |  |  |  |  | (2.8752) |  | (2.8162) |
| House price growth rate |  |  |  |  |  |  | -4.8229** | -5.0474** |
|  |  |  |  |  |  |  | (2.0293) | (1.9862) |
| Constant | 29.4963*** | 28.2867*** | 31.7738*** | 28.4476*** | 28.6446*** | 29.4989*** | 31.2270*** | 32.2401*** |
|  | (3.3349) | (3.4504) | (3.6547) | (4.2699) | (4.5902) | (4.5171) | (4.6357) | (4.5498) |
| Log likelihood | -771.9569 | -770.9151 | -770.6231 | -771.8621 | -769.4425 | -765.8957 | -766.0080 | -761.9499 |
| Number of observations | 153 | 153 | 153 | 153 | 153 | 153 | 153 | 153 |

**Panel B: Determinants of the amplitude of short-term public debt expansions in EMs**

|  | M1 | M2 | M3 | M4 | M5 | M6 | M7 | M8 |
|---|---|---|---|---|---|---|---|---|
| Credit bust |  | -6.2030 |  |  | -3.5882 | -3.7307 | -6.6144 | -7.5033 |
|  |  | (23.3531) |  |  | (22.3197) | (22.8357) | (20.7367) | (21.2188) |
| House price bust |  |  | 30.7482** |  | 29.4702** | 29.5713** | 26.6550** | 27.2556** |
|  |  |  | (13.3626) |  | (13.3920) | (13.7713) | (12.4623) | (12.7970) |
| Equity price bust |  |  |  | 19.0521 | 17.7395 | 17.6356 | 13.1350 | 12.4718 |
|  |  |  |  | (11.9676) | (11.6770) | (12.0908) | (10.9418) | (11.3449) |
| Credit growth rate |  |  |  |  |  | 0.1627 |  | 0.9942 |
|  |  |  |  |  |  | (4.1732) |  | (3.8807) |
| House price growth rate |  |  |  |  |  |  | -7.2900*** | -7.3342*** |

|  | | | | | | | | (2.4008) | (2.4297) |
|---|---|---|---|---|---|---|---|---|---|
| Constant | 52.5916*** | 52.9362*** | 47.0398*** | 46.2409*** | 41.5568*** | 41.6701*** | 50.6085*** | 51.3561*** |
|  | (4.7257) | (4.9403) | (5.1486) | (6.1343) | (6.4849) | (7.1660) | (6.7158) | (7.3810) |
| Log likelihood | -357.5845 | -357.5366 | -354.1565 | -355.9029 | -352.5580 | -352.5569 | -346.4652 | -346.4171 |
| Number of observations | 72 | 72 | 72 | 72 | 72 | 72 | 72 | 72 |

**Panel C: Determinants of the amplitude of medium-term public debt expansions in AEs**

|  | M1 | M2 | M3 | M4 | M5 | M6 | M7 | M8 |
|---|---|---|---|---|---|---|---|---|
| Credit bust |  | 42.4152 |  |  | 40.8319 | 42.8959 | 42.4679 | 45.4913 |
|  |  | (31.1317) |  |  | (31.7946) | (30.6387) | (31.9824) | (30.6047) |
| House price bust |  |  | -16.2799 |  | -13.4273 | -12.4616 | -17.2746 | -18.0918 |
|  |  |  | (21.7708) |  | (21.9975) | (21.1929) | (22.6344) | (21.6446) |
| Equity price bust |  |  |  | -1.5249 | -2.1803 | -3.2097 | -3.4812 | -5.2237 |
|  |  |  |  | (17.2709) | (17.3230) | (16.6920) | (17.4685) | (16.7175) |
| Credit growth rate |  |  |  |  |  | 19.1804** |  | 20.7409** |
|  |  |  |  |  |  | (8.3479) |  | (8.4212) |
| House price growth rate |  |  |  |  |  |  | -3.6378 | -5.3979 |
|  |  |  |  |  |  |  | (4.7130) | (4.5627) |
| Constant | 51.5526*** | 48.5229*** | 54.2659*** | 52.0246*** | 51.5488*** | 56.5383*** | 54.7335*** | 61.6700*** |
|  | (6.5521) | (6.8739) | (7.5116) | (8.5003) | (9.3752) | (9.2879) | (10.2747) | (10.2199) |
| Log likelihood | -447.6317 | -446.2857 | -447.2216 | -447.6259 | -445.9797 | -442.0622 | -445.5188 | -440.9675 |
| Number of observations | 84 | 84 | 84 | 84 | 84 | 84 | 84 | 84 |

**Panel D: Determinants of the amplitude of medium-term public debt expansions in EMs**

|  | M1 | M2 | M3 | M4 | M5 | M6 | M7 | M8 |
|---|---|---|---|---|---|---|---|---|
| Credit bust |  | 70.3570** |  |  | 67.5407* | 68.8653* | 48.8904 | 43.3970 |
|  |  | (30.4923) |  |  | (32.8067) | (34.0595) | (30.2442) | (31.9621) |
| House price bust |  |  | 27.5894 |  | 27.5894 | 27.4413 | 38.4143 | 40.4770 |
|  |  |  | (27.6771) |  | (25.2932) | (26.1153) | (22.9197) | (23.6177) |
| Equity price bust |  |  |  | 29.9347 | 9.1530 | 9.1722 | 30.8837 | 34.4510 |

|  | | | | | (34.3971) | (32.8067) | (33.8690) | (30.6637) | (31.7757) |
| --- | --- | --- | --- | --- | --- | --- | --- | --- | --- |
| Credit growth rate | | | | | | -4.8244 | | | 8.7480 |
| | | | | | | (13.1030) | | | (13.0020) |
| House price growth rate | | | | | | | | -10.1758** | -11.8624** |
| | | | | | | | | (4.6118) | (5.3374) |
| Constant | 106.6396*** | 98.1114*** | 100.7873*** | 98.4755*** | 90.1043*** | 87.8765*** | 92.4863*** | 96.9207*** |
| | (8.7506) | (8.6433) | (10.5389) | (12.8719) | (12.9125) | (14.6394) | (11.4804) | (13.4548) |
| Log likelihood | -165.1541 | -160.4146 | -164.1600 | -164.3910 | -158.9844 | -158.8133 | -153.7360 | -153.1250 |
| Number of observations | 33 | 33 | 33 | 33 | 33 | 33 | 33 | 33 |

**Panel E: Determinants of the amplitude of short-term public debt contractions in AEs**

|  | M1 | M2 | M3 | M4 | M5 | M6 | M7 | M8 |
| --- | --- | --- | --- | --- | --- | --- | --- | --- |
| Credit boom | | -0.6421 | | | -1.6237 | -2.5682 | -1.4314 | -2.4221 |
| | | (4.7128) | | | (4.6734) | (4.7533) | (4.6767) | (4.7514) |
| House price boom | | | 0.7709 | | 1.3308 | 2.0249 | 0.0024 | 0.6476 |
| | | | (3.2461) | | (3.2142) | (3.2771) | (3.4716) | (3.5132) |
| Equity price boom | | | | -5.1872** | -5.3501** | -5.5366** | -5.1533** | -5.3379** |
| | | | | (2.2751) | (2.3120) | (2.3171) | (2.3199) | (2.3225) |
| Credit growth rate | | | | | | -0.7159 | | -0.7610 |
| | | | | | | (0.6700) | | (0.6707) |
| House price growth rate | | | | | | | -0.7417 | -0.7934 |
| | | | | | | | (0.7329) | (0.7334) |
| Constant | -9.7533*** | -9.7202*** | -9.8326*** | -8.5709*** | -8.5872*** | -7.9261*** | -7.9856*** | -7.2409*** |
| | (0.8965) | (0.9326) | (0.9604) | (1.0215) | (1.1126) | (1.2724) | (1.2613) | (1.4204) |
| Log likelihood | -498.3687 | -498.3572 | -498.3338 | -495.2287 | -495.0341 | -494.3124 | -494.3864 | -493.5656 |
| Number of observations | 136 | 136 | 136 | 136 | 136 | 136 | 136 | 136 |

**Panel F: Determinants of the amplitude of short-term public debt contractions in EMs**

|  | M1 | M2 | M3 | M4 | M5 | M6 | M7 | M8 |
| --- | --- | --- | --- | --- | --- | --- | --- | --- |
| Credit boom | | 6.9437* | | | 6.2385* | 6.3504* | 6.2865* | 6.4203* |

|  | M1 | M2 | M3 | M4 | M5 | M6 | M7 | M8 |
|---|---|---|---|---|---|---|---|---|
|  |  | (3.7077) |  |  | (3.8064) | (3.8616) | (3.8396) | (3.8960) |
| House price boom |  |  | -2.4525 |  | -2.4672 | -2.4463 | -2.6657 | -2.6559 |
|  |  |  | (3.0140) |  | (2.9458) | (2.9755) | (3.0001) | (3.0294) |
| Equity price boom |  |  |  | -3.2870 | -2.4094 | -2.5356 | -2.3920 | -2.5377 |
|  |  |  |  | (2.6372) | (2.6567) | (2.7151) | (2.6791) | (2.7375) |
| Credit growth rate |  |  |  |  |  | 0.0981 |  | 0.1141 |
|  |  |  |  |  |  | (0.3246) |  | (0.3289) |
| House price growth rate |  |  |  |  |  |  | -0.2140 | -0.2296 |
|  |  |  |  |  |  |  | (0.4533) | (0.4599) |
| Constant | -9.0363*** | -9.5545*** | -8.5970*** | -8.4966*** | -8.6644*** | -8.8212*** | -8.4499*** | -8.6168*** |
|  | (0.8580) | (0.8818) | (1.0162) | (0.9569) | (1.1449) | (1.2674) | (1.2406) | (1.3418) |
| Log likelihood | -215.8616 | -213.5458 | -215.4120 | -214.8160 | -212.4825 | -212.4162 | -212.3206 | -212.2310 |
| Number of observations | 67 | 67 | 67 | 67 | 67 | 67 | 67 | 67 |

**Panel G: Determinants of the amplitude of medium-term public debt contractions in AEs**

|  | M1 | M2 | M3 | M4 | M5 | M6 | M7 | M8 |
|---|---|---|---|---|---|---|---|---|
| Credit boom |  | 12.4766 |  |  | 10.1944 | 8.2854 | 13.6002 | 11.6417 |
|  |  | (10.0661) |  |  | (10.3316) | (10.4937) | (10.6956) | (10.9063) |
| House price boom |  |  | -9.8000* |  | -6.5914 | -8.6956 | -3.6546 | -5.7685 |
|  |  |  | (5.2442) |  | (6.1469) | (6.4796) | (6.6205) | (6.9935) |
| Equity price boom |  |  |  | -5.8617 | -3.6714 | -2.4130 | -4.7945 | -3.5650 |
|  |  |  |  | (4.3617) | (4.9033) | (5.0531) | (4.9768) | (5.1489) |
| Credit growth rate |  |  |  |  |  | -2.1988 |  | -2.0406 |
|  |  |  |  |  |  | (2.1520) |  | (2.1515) |
| House price growth rate |  |  |  |  |  |  | 1.7704 | 1.6732 |
|  |  |  |  |  |  |  | (1.5214) | (1.5268) |
| Constant | -14.6508*** | -15.0125*** | -13.2305*** | -13.2066*** | -13.0865*** | -11.2119*** | -14.6665*** | -12.8400*** |
|  | (1.4511) | (1.4716) | (1.6033) | (1.7952) | (1.8362) | (2.5950) | (2.2773) | (2.9845) |
| Log likelihood | -254.1519 | -252.9408 | -251.4580 | -252.7325 | -250.3651 | -249.4762 | -249.2166 | -248.4299 |
| Number of observations | 69 | 69 | 69 | 69 | 69 | 69 | 69 | 69 |

**Panel H: Determinants of the amplitude of medium-term public debt contractions in EMs**

| | M1 | M2 | M3 | M4 | M5 | M6 | M7 | M8 |
|---|---|---|---|---|---|---|---|---|
| Credit boom | | 12.4651 | | | 15.7882* | 15.2629* | 15.7497* | 15.2818* |
| | | (9.6672) | | | (7.9599) | (7.8539) | (8.3923) | (8.3323) |
| House price boom | | | -9.3869** | | -12.2354** | -12.2595** | -11.7900* | -12.5438* |
| | | | (4.1204) | | (4.0898) | (4.0285) | (5.9323) | (5.9235) |
| Equity price boom | | | | -0.4018 | 6.6463 | 8.2879* | 6.5861 | 8.3441 |
| | | | | (4.9302) | (4.2120) | (4.3902) | (4.4710) | (4.7252) |
| Credit growth rate | | | | | | -1.2723 | | -1.2860 |
| | | | | | | (1.1129) | | (1.1966) |
| House price growth rate | | | | | | | 0.0986 | -0.0629 |
| | | | | | | | (0.9028) | (0.9077) |
| Constant | -14.2107*** | -14.6405*** | -11.9449*** | -14.0999*** | -13.6352*** | -11.7939*** | -13.8177*** | -11.6577*** |
| | (1.5026) | (1.5032) | (1.6422) | (2.0723) | (1.6874) | (2.3145) | (2.4394) | (3.1447) |
| Log likelihood | -90.1503 | -88.2688 | -84.9387 | -90.1422 | -78.9373 | -76.9712 | -78.9181 | -76.9625 |
| Number of observations | 29 | 29 | 29 | 29 | 29 | 29 | 29 | 29 |

Note: This table reports the impact of financial cycles on the amplitude of public debt cycles. All regressions control the country's fixed effect. Standard errors are in parentheses. *, ** and *** represent significance at the 10%, 5% and 1% level, respectively.

# Appendix C

Appendix C presents the results for the differences in the impact of certain financial cycles between AEs and EMs.

Table C.1 examines the differences in the impact of short-term equity price busts on the duration of public debt expansions between AEs and EMs. M1 represents the benchmark model. M2 to M5 incorporate financial variables.

Table C.1. Impact on the duration of short-term public debt expansions by Equity price bust: AEs v.s. EMs.

| | Dependent variable: duration of public debt expansions | | | | |
|---|---|---|---|---|---|
| | M1 | M2 | M3 | M4 | M5 |
| Equity price bust | | 0.2637** | 0.1492 | 0.1603 | 0.1845 |
| | | (0.1134) | (0.1220) | (0.1231) | (0.1172) |
| Equity price bust*EMs | | | 0.4036** | 0.3454* | 0.2204 |
| | | | (0.2020) | (0.1996) | (0.2017) |
| Credit bust | | | | 0.2042 | 0.0096 |
| | | | | (0.2193) | (0.2186) |
| House price bust | | | | 0.3453** | 0.3449*** |
| | | | | (0.1366) | (0.1314) |
| Credit growth rate | | | | | 0.0560 |
| | | | | | (0.0392) |
| House price growth rate | | | | | -0.0853*** |
| | | | | | (0.0227) |
| Constant | 2.1544*** | 2.0674*** | 2.0638*** | 1.9933*** | 2.0615*** |
| | (0.0746) | (0.0774) | (0.0748) | (0.0778) | (0.0822) |
| Weibull shape parameter (ln $p$) | 0.3610*** | 0.3557*** | 0.3586*** | 0.3662*** | 0.4253*** |
| | (0.0589) | (0.0584) | (0.0576) | (0.0590) | (0.0615) |
| Frailty parameter (ln $\theta$) | -1.3592*** | -1.7169*** | -1.9210*** | -2.0128*** | -1.5181*** |
| | (0.5147) | (0.5945) | (0.6233) | (0.6790) | (0.5804) |
| Log likelihood | -278.9740 | -276.2593 | -274.2519 | -270.4956 | -263.0025 |
| LR Chi-squared | | 5.4295** | 9.4442*** | 16.9567*** | 31.9430*** |
| Number of observations | 225 | 225 | 225 | 225 | 225 |

Note: This table reports the impact of financial cycles on the duration of public debt cycles. All regressions control the shared frailty. Standard errors are in parentheses. $p$ is the Weibull shape parameter, reflecting duration time-dependency. $\theta$ is the variance of $\alpha_i$ in Eq. (4). Equity price bust*EMs is the interaction term of Equity price bust dummy and dummy that refers to EMs. *, ** and *** represent significance at the 10%, 5% and 1% level,

respectively.

Table C.2 presents the results examining the differences in the impact of short-term equity price booms on the duration of public debt contractions between AEs and EMs. M1 is the benchmark model. M2 to M5 incorporate financial variables.

**Table C.2. Impact on the duration of short-term public debt contractions by Equity price boom: AEs v.s. EMs.**

| | \multicolumn{5}{c}{Dependent variable: duration of public debt contractions} | | | | |
|---|---|---|---|---|---|
| | M1 | M2 | M3 | M4 | M5 |
| Equity price boom | | 0.5223*** | 0.5449*** | 0.5696*** | 0.5743*** |
| | | (0.1353) | (0.1533) | (0.1517) | (0.1518) |
| Equity price boom*EMs | | | -0.0868 | -0.0834 | 0.0068 |
| | | | (0.2705) | (0.2648) | (0.2663) |
| Credit boom | | | | 0.0988 | 0.0706 |
| | | | | (0.2273) | (0.2266) |
| House price boom | | | | -0.1560 | -0.1101 |
| | | | | (0.1666) | (0.1753) |
| Credit growth rate | | | | | -0.0394* |
| | | | | | (0.0227) |
| House price growth rate | | | | | 0.0417 |
| | | | | | (0.0323) |
| Constant | 1.9734*** | 1.8761*** | 1.8767*** | 1.8937*** | 1.8933*** |
| | (0.0727) | (0.0673) | (0.0672) | (0.0702) | (0.0835) |
| Weibull shape parameter (ln $p$) | 0.2948*** | 0.3053*** | 0.3048*** | 0.2959*** | 0.3043*** |
| | (0.0630) | (0.0622) | (0.0624) | (0.0639) | (0.0636) |
| Frailty parameter (ln $\theta$) | -2.1438*** | -3.3716 | -3.4220 | -4.8702 | -4.7992 |
| | (0.8179) | (2.0937) | (2.1974) | (8.7846) | (7.8335) |
| Log likelihood | -254.8353 | -247.3315 | -247.2810 | -246.7623 | -244.7356 |
| LR Chi-squared | | 15.0074**** | 15.1086*** | 16.1460*** | 20.1994*** |
| Number of observations | 203 | 203 | 203 | 203 | 203 |

Note: This table reports the robustness tests for the impact of financial cycles on the duration of public debt cycles. All regressions control the shared frailty. Standard errors are in parentheses. $p$ is the Weibull shape parameter, reflecting duration time-dependency. $\theta$ is the variance of $\alpha_i$ in Eq. (4). Equity price boom*EMs is the interaction term of Equity price boom dummy and dummy that refers to EMs. *, ** and *** represent significance at the 10%, 5% and 1% level, respectively.

Table C.3 presents the regression results examining the differences in the impact of medium-term house price booms on the duration of public debt contractions between AEs and EMs. M1 is the benchmark model. M2 to M5 incorporate financial variables.

**Table C.3. Impact on the duration of medium-term public debt contractions by House price boom: AEs v.s. EMs.**

| | Dependent variable: duration of public debt contractions | | | | |
|---|---|---|---|---|---|
| | M1 | M2 | M3 | M4 | M5 |
| House price boom | | 0.4962*** | 0.4515** | 0.4006* | 0.3477 |
| | | (0.1659) | (0.2156) | (0.2311) | (0.2342) |
| House price boom*EMs | | | 0.0999 | 0.1371 | 0.1955 |
| | | | (0.3152) | (0.3189) | (0.3205) |
| Credit boom | | | | -0.2764 | -0.3167 |
| | | | | (0.3482) | (0.3499) |
| Equity price boom | | | | 0.0361 | 0.0977 |
| | | | | (0.1541) | (0.1632) |
| Credit growth rate | | | | | -0.0520 |
| | | | | | (0.0420) |
| Constant | 2.5485*** | 2.4204*** | 2.4202*** | 2.4188*** | 2.4747*** |
| | (0.1034) | (0.1116) | (0.1113) | (0.1135) | (0.1181) |
| Weibull shape parameter (ln $p$) | 0.5694*** | 0.6673*** | 0.6660*** | 0.6725*** | 0.6733*** |
| | (0.1032) | (0.0989) | (0.0985) | (0.0994) | (0.0987) |
| Frailty parameter (ln $\theta$) | -0.8572 | -0.3801 | -0.3877 | -0.3475 | -0.4162 |
| | (0.7897) | (0.6726) | (0.6692) | (0.6782) | (0.6619) |
| Log likelihood | -106.4406 | -101.9877 | -101.9372 | -101.6277 | -100.9394 |
| LR Chi-squared | | 8.9059*** | 9.0068*** | 9.6259** | 11.0025** |
| Number of observations | 98 | 98 | 98 | 98 | 98 |

Note: This table reports the robustness tests for the impact of financial cycles on the duration of public debt cycles. All regressions control the shared frailty. Standard errors are in parentheses. $p$ is the Weibull shape parameter, reflecting duration time-dependency. $\theta$ is the variance of $\alpha_i$ in Eq. (4). House price boom*EMs is the interaction term of House price boom dummy and dummy that refers to EMs. The house price growth rate is dropped because it causes significant multi-collinearity in the medium-term contraction regressions. *, ** and *** represent significance at the 10%, 5% and 1% level, respectively.

# Appendix D

Appendix D summarizes the results for robustness tests (incorporating macroeconomic variables and orthogonal robustness tests).

Table D.1 presents the results on the duration of public debt cycles including macroeconomic variables. Columns (1) to (6) are the regressions with different macroeconomic variables added one by one into the main model, on the basis of the three dummy variables and growth rate variables.

**Table D.1. Robustness test (with macroeconomic variables): determinants of the duration of public debt cycles.**

**Panel A: Robustness test: determinants of the duration of short-term public debt expansions in AEs**

|  | (1) | (2) | (3) | (4) | (5) | (6) |
|---|---|---|---|---|---|---|
| Credit bust | -0.1176 | -0.0977 | -0.1674 | -0.0966 | 0.0295 | -0.1388 |
|  | (0.2533) | (0.2562) | (0.2506) | (0.2542) | (0.2640) | (0.2564) |
| House price bust | 0.1600 | 0.2085 | 0.1563 | 0.1647 | 0.1178 | 0.1778 |
|  | (0.1543) | (0.1473) | (0.1515) | (0.1526) | (0.1517) | (0.1533) |
| Equity price bust | 0.2244* | 0.1943* | 0.2127* | 0.2294* | 0.2261** | 0.1982* |
|  | (0.1171) | (0.1116) | (0.1141) | (0.1174) | (0.1152) | (0.1186) |
| Credit growth rate | 0.0784* | 0.0655 | 0.0728 | 0.0799* | 0.0547 | 0.0780* |
|  | (0.0454) | (0.0407) | (0.0465) | (0.0456) | (0.0444) | (0.0458) |
| House price growth rate | -0.1101*** | -0.1263*** | -0.1331*** | -0.1044*** | -0.0638* | -0.1082*** |
|  | (0.0281) | (0.0286) | (0.0292) | (0.0280) | (0.0326) | (0.0280) |
| GDP | 0.0329 |  |  |  |  |  |
|  | (0.0579) |  |  |  |  |  |
| Monetary |  | 0.1503*** |  |  |  |  |
|  |  | (0.0370) |  |  |  |  |
| CPI |  |  | 0.3150*** |  |  |  |
|  |  |  | (0.1179) |  |  |  |
| REER |  |  |  | -0.0247 |  |  |
|  |  |  |  | (0.0353) |  |  |
| Balance |  |  |  |  | -0.0336*** |  |
|  |  |  |  |  | (0.0111) |  |
| Oil |  |  |  |  |  | 0.0078 |
|  |  |  |  |  |  | (0.0082) |
| Constant | 2.0299*** | 1.8131*** | 1.9156*** | 2.0334*** | 2.0712*** | 2.0096*** |
|  | (0.1095) | (0.1136) | (0.1110) | (0.1092) | (0.1040) | (0.1123) |

| | (1) | (2) | (3) | (4) | (5) | (6) |
|---|---|---|---|---|---|---|
| Weibull shape parameter (ln $p$) | 0.5165*** | 0.5706*** | 0.5318*** | 0.5187*** | 0.5397*** | 0.5183*** |
| | (0.0735) | (0.0699) | (0.0724) | (0.0734) | (0.0723) | (0.0733) |
| Frailty parameter (ln $\theta$) | -0.7305 | -0.7390 | -0.9167 | -0.7237 | -0.7870 | -0.7270 |
| | (0.5661) | (0.5751) | (0.5910) | (0.5627) | (0.5947) | (0.5668) |
| Log likelihood | -171.8023 | -164.0969 | -168.3927 | -171.7161 | -167.5730 | -171.5055 |
| LR Chi-squared | 18.3285*** | 33.7394*** | 25.1477*** | 18.5009*** | 26.7872*** | 18.9221*** |
| Number of observations | 153 | 153 | 153 | 153 | 153 | 153 |

**Panel B: Robustness test: determinants of the duration of short-term public debt expansions in EMs**

| | (1) | (2) | (3) | (4) | (5) | (6) |
|---|---|---|---|---|---|---|
| Credit bust | -0.0015 | -0.0123 | 0.0376 | 0.0836 | -0.0104 | 0.0574 |
| | (0.3856) | (0.3856) | (0.3832) | (0.3888) | (0.3888) | (0.3836) |
| House price bust | 0.4897** | 0.4885** | 0.5086** | 0.4182* | 0.4929** | 0.4564** |
| | (0.2451) | (0.2312) | (0.2303) | (0.2338) | (0.2350) | (0.2288) |
| Equity price bust | 0.4356** | 0.4220** | 0.4765** | 0.4851** | 0.4248** | 0.4921** |
| | (0.2014) | (0.2032) | (0.2025) | (0.2014) | (0.2097) | (0.2063) |
| Credit growth rate | 0.0259 | 0.0278 | 0.0034 | -0.0134 | 0.0232 | 0.0291 |
| | (0.0796) | (0.0723) | (0.0732) | (0.0768) | (0.0716) | (0.0700) |
| House price growth rate | -0.0725** | -0.0746** | -0.0667* | -0.0618* | -0.0722** | -0.0676** |
| | (0.0346) | (0.0346) | (0.0350) | (0.0342) | (0.0344) | (0.0345) |
| GDP | 0.0026 | | | | | |
| | (0.0440) | | | | | |
| Monetary | | 0.0163 | | | | |
| | | (0.0368) | | | | |
| CPI | | | -0.0449 | | | |
| | | | (0.0399) | | | |
| REER | | | | 0.0485 | | |
| | | | | (0.0393) | | |
| Balance | | | | | -0.0036 | |
| | | | | | (0.0175) | |
| Oil | | | | | | 0.0106 |
| | | | | | | (0.0094) |
| Constant | 2.0676*** | 2.0379*** | 2.0885*** | 2.0107*** | 2.0722*** | 2.0370*** |
| | (0.1438) | (0.1561) | (0.1379) | (0.1446) | (0.1407) | (0.1410) |
| Weibull shape parameter (ln $p$) | 0.3111*** | 0.3126*** | 0.3204*** | 0.3234*** | 0.3104*** | 0.3262*** |
| | (0.0900) | (0.0898) | (0.0901) | (0.0908) | (0.0899) | (0.0913) |
| Frailty parameter (ln $\theta$) | -15.5386 | -16.3041 | -15.2585 | -14.3370 | -14.9272 | -15.1933 |
| | (1187.1711) | (1560.5607) | (1028.2094) | (806.5806) | (920.7045) | (1035.6727) |
| Log likelihood | -87.1404 | -87.0402 | -86.6223 | -86.4100 | -87.1214 | -86.5267 |
| LR Chi-squared | 17.8408*** | 18.0413*** | 18.8769*** | 19.3016*** | 17.8788*** | 19.0682*** |
| Number of observations | 72 | 72 | 72 | 72 | 72 | 72 |

**Panel C: Robustness test: determinants of the duration of medium-term public debt expansions in AEs**

| | (1) | (2) | (3) | (4) | (5) | (6) |
|---|---|---|---|---|---|---|
| Credit bust | 0.1328 | 0.0818 | 0.1168 | 0.1152 | 0.1503 | 0.0723 |
| | (0.2799) | (0.2548) | (0.2715) | (0.2743) | (0.2740) | (0.2670) |
| House price bust | 0.1782 | 0.1409 | 0.1796 | 0.1907 | 0.1592 | 0.1441 |
| | (0.1778) | (0.1665) | (0.1775) | (0.1834) | (0.1769) | (0.1729) |
| Equity price bust | 0.1334 | 0.1142 | 0.1029 | 0.1348 | 0.1406 | 0.1251 |
| | (0.1420) | (0.1327) | (0.1457) | (0.1420) | (0.1411) | (0.1402) |
| Credit growth rate | 0.2484*** | 0.1818*** | 0.2548*** | 0.2476*** | 0.2207*** | 0.2521*** |
| | (0.0617) | (0.0657) | (0.0608) | (0.0624) | (0.0654) | (0.0598) |
| House price growth rate | -0.0251 | -0.0506* | -0.0375 | -0.0221 | 0.0053 | -0.0177 |
| | (0.0250) | (0.0276) | (0.0278) | (0.0236) | (0.0304) | (0.0238) |
| GDP | 0.0338 | | | | | |
| | (0.0882) | | | | | |
| Monetary | | 0.2011*** | | | | |
| | | (0.0632) | | | | |
| CPI | | | 0.1385 | | | |
| | | | (0.1262) | | | |
| REER | | | | -0.0002 | | |
| | | | | (0.0008) | | |
| Balance | | | | | -0.0180 | |
| | | | | | (0.0129) | |
| Oil | | | | | | 0.0163** |
| | | | | | | (0.0083) |
| Constant | 2.6626*** | 2.3155*** | 2.6351*** | 2.6711*** | 2.6874*** | 2.5968*** |
| | (0.1097) | (0.1546) | (0.1091) | (0.1065) | (0.1031) | (0.1089) |
| Weibull shape parameter (ln $p$) | 0.6291*** | 0.7204*** | 0.6038*** | 0.6271*** | 0.6218*** | 0.6399*** |
| | (0.1062) | (0.1026) | (0.1066) | (0.1076) | (0.1118) | (0.1010) |
| Frailty parameter (ln $\theta$) | -1.7156 | -1.0599 | -2.2040 | -1.7586 | -2.0020 | -1.8327* |
| | (1.1399) | (0.7935) | (1.5328) | (1.1761) | (1.4028) | (1.1034) |
| Log likelihood | -81.9679 | -77.1299 | -81.4187 | -82.0125 | -81.0952 | -79.9876 |
| LR Chi-squared | 18.2460*** | 27.9220*** | 19.3445*** | 18.1567*** | 19.9915*** | 22.2066*** |
| Number of observations | 84 | 84 | 84 | 84 | 84 | 84 |

**Panel D: Robustness test: determinants of the duration of medium-term public debt expansions in EMs**

|  | (1) | (2) | (3) | (4) | (5) | (6) |
|---|---|---|---|---|---|---|
| Credit bust | 0.6057* | 0.6779* | 0.8383** | 0.7755** | 0.6998* | 0.6623* |
|  | (0.3682) | (0.3677) | (0.3543) | (0.3573) | (0.3857) | (0.3565) |
| House price bust | 0.5598* | 0.4633* | 0.7237*** | 0.4551* | 0.4426* | 0.4842* |
|  | (0.2887) | (0.2625) | (0.2707) | (0.2716) | (0.2680) | (0.2653) |
| Equity price bust | -0.0094 | 0.0969 | 0.1451 | -0.0250 | 0.0243 | 0.0653 |
|  | (0.2575) | (0.2647) | (0.2498) | (0.2723) | (0.2543) | (0.2636) |
| Credit growth rate | 0.1070 | -0.0129 | 0.0185 | 0.0148 | 0.0580 | 0.0585 |
|  | (0.1077) | (0.1102) | (0.0825) | (0.0957) | (0.0862) | (0.0875) |
| House price growth rate | -0.0456 | -0.0319 | -0.0410 | -0.0207 | -0.0214 | -0.0254 |
|  | (0.0498) | (0.0503) | (0.0489) | (0.0474) | (0.0523) | (0.0492) |
| GDP | 0.0421 |  |  |  |  |  |
|  | (0.0469) |  |  |  |  |  |
| Monetary |  | -0.0305 |  |  |  |  |
|  |  | (0.0335) |  |  |  |  |
| CPI |  |  | -0.1654** |  |  |  |
|  |  |  | (0.0746) |  |  |  |
| REER |  |  |  | 0.0594 |  |  |
|  |  |  |  | (0.0512) |  |  |
| Balance |  |  |  |  | 0.0106 |  |
|  |  |  |  |  | (0.0182) |  |
| Oil |  |  |  |  |  | 0.0131 |
|  |  |  |  |  |  | (0.0145) |
| Constant | 2.7429*** | 2.7635*** | 2.7853*** | 2.6195*** | 2.7340*** | 2.6611*** |
|  | (0.2039) | (0.1974) | (0.1721) | (0.2146) | (0.1973) | (0.2177) |
| Weibull shape parameter | 0.6178*** | 0.6382*** | 0.6976*** | 0.7343*** | 0.6164*** | 0.6639*** |
| (ln $p$) | (0.1814) | (0.1748) | (0.1655) | (0.1798) | (0.1781) | (0.1846) |
| Frailty parameter (ln $\theta$) | -2.3536 | -1.6818 | -1.5548 | -0.4088 | -1.9619 | -1.2772 |
|  | (3.6236) | (2.1406) | (1.5772) | (1.2556) | (2.5480) | (1.7507) |
| Log likelihood | -32.5185 | -32.4743 | -30.6549 | -32.4411 | -32.7095 | -32.4888 |
| LR Chi-squared | 9.9566 | 10.0449 | 13.6838** | 10.1115 | 9.5747 | 10.0160 |
| Number of observations | 33 | 33 | 33 | 33 | 33 | 33 |

**Panel E: Robustness test: determinants of the duration of short-term public debt contractions in AEs**

|  | (1) | (2) | (3) | (4) | (5) | (6) |
|---|---|---|---|---|---|---|
| Credit boom | 0.4777 | 0.4793 | 0.4835 | 0.5925** | 0.4873 | 0.4645 |
|  | (0.2945) | (0.2944) | (0.2930) | (0.2784) | (0.2915) | (0.2913) |
| House price boom | 0.1614 | 0.1575 | 0.1432 | 0.2065 | 0.1629 | 0.1086 |

|  | (1) | (2) | (3) | (4) | (5) | (6) |
|---|---|---|---|---|---|---|
|  | (0.2278) | (0.2260) | (0.2276) | (0.2095) | (0.2251) | (0.2241) |
| Equity price boom | 0.5979*** | 0.5995*** | 0.6029*** | 0.6211*** | 0.5991*** | 0.5831*** |
|  | (0.1694) | (0.1693) | (0.1683) | (0.1424) | (0.1641) | (0.1732) |
| Credit growth rate | 0.0034 | 0.0029 | 0.0030 | -0.0217 | 0.0008 | -0.0046 |
|  | (0.0640) | (0.0645) | (0.0644) | (0.0576) | (0.0636) | (0.0645) |
| House price growth rate | 0.0957** | 0.0950* | 0.0971** | 0.1364*** | 0.0945** | 0.0859* |
|  | (0.0486) | (0.0485) | (0.0483) | (0.0478) | (0.0478) | (0.0484) |
| GDP | -0.0053 |  |  |  |  |  |
|  | (0.0769) |  |  |  |  |  |
| Monetary |  | 0.0033 |  |  |  |  |
|  |  | (0.0611) |  |  |  |  |
| CPI |  |  | 0.0579 |  |  |  |
|  |  |  | (0.1521) |  |  |  |
| REER |  |  |  | -0.1001** |  |  |
|  |  |  |  | (0.0402) |  |  |
| Balance |  |  |  |  | -0.0058 |  |
|  |  |  |  |  | (0.0106) |  |
| Oil |  |  |  |  |  | -0.0063 |
|  |  |  |  |  |  | (0.0063) |
| Constant | 1.7592*** | 1.7519*** | 1.7290*** | 1.7168*** | 1.7779*** | 1.7905*** |
|  | (0.1228) | (0.1361) | (0.1325) | (0.1032) | (0.1176) | (0.1168) |
| Weibull shape parameter (ln $p$) | 0.3421*** | 0.3411*** | 0.3393*** | 0.3704*** | 0.3331*** | 0.3366*** |
|  | (0.0773) | (0.0779) | (0.0780) | (0.0629) | (0.0763) | (0.0789) |
| Frailty parameter (ln $\theta$) | -3.9184 | -4.0029 | -4.3583 | -16.7911 | -9.8084 | -7.2688 |
|  | (4.1159) | (4.5545) | (6.3681) | (1631.9644) | (1292.9261) | (110.4474) |
| Log likelihood | -160.3358 | -160.3367 | -160.2652 | -157.4810 | -160.2015 | -159.8402 |
| LR Chi-squared | 18.1747*** | 18.1729*** | 18.3160*** | 23.8844*** | 18.4433*** | 19.1659*** |
| Number of observations | 136 | 136 | 136 | 136 | 136 | 136 |

**Panel F: Robustness test: determinants of the duration of short-term public debt contractions in EMs**

|  | (1) | (2) | (3) | (4) | (5) | (6) |
|---|---|---|---|---|---|---|
| Credit boom | -0.8941*** | -0.8991*** | -0.8880*** | -0.8749*** | -0.9314*** | -0.9410*** |
|  | (0.3324) | (0.3315) | (0.3355) | (0.3312) | (0.3317) | (0.3390) |
| House price boom | -0.3990* | -0.4125* | -0.4434* | -0.4229* | -0.4042* | -0.4051* |
|  | (0.2417) | (0.2353) | (0.2389) | (0.2318) | (0.2346) | (0.2369) |
| Equity price boom | 0.5807** | 0.5282** | 0.5103** | 0.5941** | 0.5046** | 0.4794* |
|  | (0.2726) | (0.2396) | (0.2431) | (0.2563) | (0.2388) | (0.2496) |
| Credit growth rate | -0.0593** | -0.0315 | -0.0425 | -0.0592** | -0.0565** | -0.0507** |
|  | (0.0288) | (0.0387) | (0.0304) | (0.0256) | (0.0235) | (0.0227) |
| House price growth rate | 0.0006 | 0.0013 | -0.0019 | -0.0012 | -0.0074 | 0.0039 |
| | (0.0383) | (0.0385) | (0.0386) | (0.0379) | (0.0398) | (0.0387) |
| GDP | 0.0207 | | | | | |
| | (0.0413) | | | | | |
| Monetary | | 0.0205 | | | | |
| | | (0.0349) | | | | |
| CPI | | | -0.0478 | | | |
| | | | (0.0925) | | | |
| REER | | | | 0.0202 | | |
| | | | | (0.0258) | | |
| Balance | | | | | -0.0193 | |
| | | | | | (0.0233) | |
| Oil | | | | | | -0.0050 |
| | | | | | | (0.0094) |
| Constant | 2.0416*** | 2.0168*** | 2.1297*** | 2.0873*** | 2.0998*** | 2.1167*** |
| | (0.1545) | (0.1709) | (0.1636) | (0.1294) | (0.1307) | (0.1431) |
| Weibull shape parameter (ln $p$) | 0.3589*** | 0.3603*** | 0.3503*** | 0.3670*** | 0.3652*** | 0.3627*** |
| | (0.0882) | (0.0881) | (0.0891) | (0.0901) | (0.0886) | (0.0891) |
| Frailty parameter (ln $\theta$) | -15.1445 | -16.4992 | -15.9231 | -14.2928 | -15.2476 | -15.6361 |
| | (2219.4435) | (1789.4073) | (2443.5415) | (885.2398) | (865.3449) | (1139.4737) |
| Log likelihood | -78.6287 | -78.5787 | -77.9736 | -78.4560 | -78.4279 | -78.6102 |
| LR Chi-squared | 12.9835** | 13.0837** | 12.3841* | 13.3291** | 13.3852** | 13.0206** |
| Number of observations | 67 | 67 | 66 | 67 | 67 | 67 |

**Panel G: Robustness test: determinants of the duration of medium-term public debt contractions in AEs**

| | (1) | (2) | (3) | (4) | (5) | (6) |
|---|---|---|---|---|---|---|
| Credit boom | 0.0051 | 0.0414 | -0.0461 | 0.0107 | 0.0112 | -0.1034 |
| | (0.4179) | (0.4166) | (0.4081) | (0.4143) | (0.4210) | (0.3955) |
| House price boom | 0.4506* | 0.4471* | 0.3819 | 0.5170** | 0.4495* | 0.4490* |
| | (0.2518) | (0.2414) | (0.2480) | (0.2411) | (0.2475) | (0.2373) |
| Equity price boom | 0.0651 | 0.0722 | 0.0719 | 0.0157 | 0.0695 | 0.1971 |
| | (0.1969) | (0.1920) | (0.1967) | (0.1944) | (0.1948) | (0.1913) |
| Credit growth rate | 0.0120 | 0.0121 | 0.0278 | 0.0362 | -0.0008 | 0.0340 |
| | (0.0939) | (0.0930) | (0.0951) | (0.0946) | (0.0999) | (0.0895) |
| GDP | -0.0329 | | | | | |
| | (0.0992) | | | | | |
| Monetary | | 0.0614 | | | | |
| | | (0.0676) | | | | |
| CPI | | | 0.1779 | | | |

|  | (1) | (2) | (3) | (4) | (5) | (6) |
|---|---|---|---|---|---|---|
|  |  |  |  | (0.1809) |  |  |
| REER |  |  |  |  | 0.0157 |  |
|  |  |  |  |  | (0.0096) |  |
| Balance |  |  |  |  |  | -0.0070 |
|  |  |  |  |  |  | (0.0163) |
| Oil |  |  |  |  |  | -0.0186*** |
|  |  |  |  |  |  | (0.0070) |
| Constant | 2.4202*** | 2.2886*** | 2.2708*** | 2.4212*** | 2.4399*** | 2.3202*** |
|  | (0.1737) | (0.1903) | (0.2068) | (0.1515) | (0.1857) | (0.1532) |
| Weibull shape parameter | 0.6547*** | 0.6647*** | 0.6981*** | 0.6622*** | 0.6412*** | 0.7245*** |
| ($\ln p$) | (0.1168) | (0.1159) | (0.1217) | (0.1191) | (0.1223) | (0.1185) |
| Frailty parameter ($\ln \theta$) | -0.6837 | -0.6012 | -0.2840 | -0.7709 | -0.8774 | -0.3554 |
|  | (0.8446) | (0.8272) | (0.8908) | (0.8753) | (1.0474) | (0.7553) |
| Log likelihood | -70.7320 | -70.3705 | -70.3326 | -69.6951 | -70.7006 | -67.1795 |
| LR Chi-squared | 4.9456 | 5.6685 | 5.7443 | 7.0192 | 5.0082 | 12.0506** |
| Number of observations | 69 | 69 | 69 | 69 | 69 | 69 |

**Panel H: Robustness test: determinants of the duration of medium-term public debt contractions in EMs**

|  | (1) | (2) | (3) | (4) | (5) | (6) |
|---|---|---|---|---|---|---|
| Credit boom | -1.4395*** | -1.4060*** | -1.4270*** | -1.4641*** | -1.4535*** | -1.4587*** |
|  | (0.4787) | (0.4701) | (0.4728) | (0.4862) | (0.4914) | (0.4666) |
| House price boom | 0.5730** | 0.5420** | 0.5759** | 0.5776** | 0.5830** | 0.5961** |
|  | (0.2461) | (0.2376) | (0.2457) | (0.2521) | (0.2478) | (0.2462) |
| Equity price boom | 0.0403 | 0.1484 | 0.0339 | 0.0202 | 0.0262 | 0.0557 |
|  | (0.2875) | (0.2866) | (0.2967) | (0.2910) | (0.2835) | (0.3011) |
| Credit growth rate | -0.0754 | -0.0354 | -0.0754 | -0.0740 | -0.0740 | -0.0750 |
|  | (0.0483) | (0.0561) | (0.0848) | (0.0474) | (0.0499) | (0.0476) |
| GDP | 0.0090 |  |  |  |  |  |
|  | (0.0412) |  |  |  |  |  |
| Monetary |  | 0.0542 |  |  |  |  |
|  |  | (0.0455) |  |  |  |  |
| CPI |  |  | 0.0511 |  |  |  |
|  |  |  | (0.1100) |  |  |  |
| REER |  |  |  | -0.0021 |  |  |
|  |  |  |  | (0.0255) |  |  |
| Balance |  |  |  |  | -0.0002 |  |
|  |  |  |  |  | (0.0351) |  |
| Oil |  |  |  |  |  | 0.0029 |
|  |  |  |  |  |  | (0.0098) |
| Constant | 2.4033*** | 2.2863*** | 2.3194*** | 2.3661*** | 2.3669*** | 2.3422*** |
| | (0.3366) | (0.3319) | (0.3726) | (0.3656) | (0.3645) | (0.3698) |
| Weibull shape parameter ($\ln p$) | 0.9196*** | 0.9335*** | 0.9321*** | 0.9431*** | 0.9414*** | 0.9446*** |
| | (0.1760) | (0.1838) | (0.1711) | (0.1678) | (0.1676) | (0.1636) |
| Frailty parameter ($\ln \theta$) | 1.1492 | 0.9775 | 1.3385 | 1.4137 | 1.3975 | 1.4311 |
| | (1.5905) | (1.5767) | (1.6479) | (1.5595) | (1.5483) | (1.5314) |
| Log likelihood | -27.4275 | -26.7577 | -26.6416 | -27.4444 | -27.4478 | -27.4053 |
| LR Chi-squared | 10.3902* | 11.7298** | 8.2682 | 10.3565* | 10.3497* | 10.4346* |
| Number of observations | 29 | 29 | 28 | 29 | 29 | 29 |

Note: This table reports the robustness tests for the impact of financial cycles on the duration of public debt cycles. All regressions control the shared frailty. Standard errors are in parentheses. $p$ is the Weibull shape parameter, reflecting duration time-dependency. $\theta$ is the variance of $\alpha_i$ in Eq. (4). The house price growth rate is dropped from medium-term contraction regressions because it causes significant multi-collinearity. *, ** and *** represent significance at the 10%, 5% and 1% level, respectively.

Table D.2 presents the results of orthogonal robustness tests for the impact of financial cycles on the duration of public debt cycles. (1) is the baseline result that only controls the shared frailty. (2) to (5) are the results including financial variables (i.e. dummy variables).

Table D.2. Orthogonal robustness test: determinants of the duration of public debt cycles.

**Panel A: Robustness test: determinants of the duration of short-term public debt expansions in AEs**

|  | (1) | (2) | (3) | (4) | (5) |
|---|---|---|---|---|---|
| Credit bust$^\perp$ |  | 0.0243 |  |  | 0.1031 |
|  |  | (0.2640) |  |  | (0.2576) |
| House price bust$^\perp$ |  |  | 0.1831 |  | 0.1653 |
|  |  |  | (0.1655) |  | (0.1651) |
| Equity price bust$^\perp$ |  |  |  | 0.2343* | 0.2301* |
|  |  |  |  | (0.1277) | (0.1287) |
| Constant | 2.1041*** | 2.1045*** | 2.1023*** | 2.1006*** | 2.1016*** |
|  | (0.0931) | (0.0931) | (0.0918) | (0.0899) | (0.0882) |
| Weibull shape parameter (ln $p$) | 0.4282*** | 0.4281*** | 0.4263*** | 0.4337*** | 0.4312*** |
|  | (0.0699) | (0.0700) | (0.0714) | (0.0696) | (0.0711) |
| Frailty parameter (ln $\theta$) | -1.1489** | -1.1513** | -1.2107** | -1.2506** | -1.3381** |
|  | (0.5810) | (0.5823) | (0.6025) | (0.5872) | (0.6185) |
| Log likelihood | -180.9666 | -180.9623 | -180.3247 | -179.2642 | -178.6956 |
| LR Chi-squared |  | 0.0085 | 1.2837 | 3.4047* | 4.5420 |
| Number of observations | 153 | 153 | 153 | 153 | 153 |

**Panel B: Robustness test: determinants of the duration of short-term public debt expansions in EMs**

|  | (1) | (2) | (3) | (4) | (5) |
|---|---|---|---|---|---|
| Credit bust$^\perp$ |  | -0.2405 |  |  | -0.0458 |
|  |  | (0.4193) |  |  | (0.3803) |
| House price bust$^\perp$ |  |  | 0.5176** |  | 0.3950* |
|  |  |  | (0.2386) |  | (0.2286) |
| Equity price bust$^\perp$ |  |  |  | 0.6115*** | 0.5561*** |
|  |  |  |  | (0.1999) | (0.2003) |
| Constant | 2.2483*** | 2.2510*** | 2.2213*** | 2.2529*** | 2.2384*** |
|  | (0.1202) | (0.1182) | (0.1206) | (0.0973) | (0.0963) |
| Weibull shape parameter (ln $p$) | 0.2220** | 0.2211** | 0.2574** | 0.2527*** | 0.2627*** |
|  | (0.1085) | (0.1082) | (0.1057) | (0.0906) | (0.0882) |
| Frailty parameter (ln $\theta$) | -2.2822 | -2.4073 | -2.0659* | -15.9505 | -15.4714 |
|  | (1.4864) | (1.6090) | (1.2538) | (992.2904) | (1220.1901) |
| Log likelihood | -96.0608 | -95.9078 | -93.4923 | -92.1026 | -90.4105 |
| LR Chi-squared | | 0.3060 | 5.1370** | 7.9165*** | 11.3006*** |
| Number of observations | 72 | 72 | 72 | 72 | 72 |

**Panel C: Robustness test: determinants of the duration of medium-term public debt expansions in AEs**

| | (1) | (2) | (3) | (4) | (5) |
|---|---|---|---|---|---|
| Credit bust$^\perp$ | | 0.0896 | | | 0.1553 |
| | | (0.2874) | | | (0.2872) |
| House price bust$^\perp$ | | | 0.1987 | | 0.1967 |
| | | | (0.1899) | | (0.1910) |
| Equity price bust$^\perp$ | | | | 0.1203 | 0.1117 |
| | | | | (0.1545) | (0.1551) |
| Constant | 2.6775*** | 2.6763*** | 2.6812*** | 2.6791*** | 2.6795*** |
| | (0.1021) | (0.1024) | (0.0984) | (0.0998) | (0.0971) |
| Weibull shape parameter (ln $p$) | 0.5434*** | 0.5465*** | 0.5373*** | 0.5406*** | 0.5423*** |
| | (0.1045) | (0.1047) | (0.1091) | (0.1052) | (0.1100) |
| Frailty parameter (ln $\theta$) | -1.1882 | -1.1575 | -1.3494 | -1.2668 | -1.3437 |
| | (0.8738) | (0.8647) | (0.9284) | (0.8817) | (0.8978) |
| Log likelihood | -91.0909 | -91.0412 | -90.5280 | -90.7850 | -90.1541 |
| LR Chi-squared | | 0.0994 | 1.1258 | 0.6118 | 1.8736 |
| Number of observations | 84 | 84 | 84 | 84 | 84 |

**Panel D: Robustness test: determinants of the duration of medium-term public debt expansions in EMs**

| | (1) | (2) | (3) | (4) | (5) |
|---|---|---|---|---|---|
| Credit bust$^\perp$ | | 0.7090* | | | 0.7124* |
| | | (0.3715) | | | (0.3652) |
| House price bust$^\perp$ | | | 0.5401* | | 0.5052* |
| | | | (0.2921) | | (0.2719) |
| Equity price bust$^\perp$ | | | | 0.1012 | 0.0053 |
| | | | | (0.2930) | (0.2611) |
| Constant | 2.9496*** | 2.8941*** | 2.9222*** | 2.9425*** | 2.8588*** |
| | (0.1330) | (0.1546) | (0.1285) | (0.1357) | (0.1435) |
| Weibull shape parameter (ln $p$) | 0.4440*** | 0.5670*** | 0.4909*** | 0.4541*** | 0.6295*** |
| | (0.1675) | (0.2008) | (0.1653) | (0.1693) | (0.1757) |
| Frailty parameter (ln $\theta$) | -3.2056 | -1.6326 | -3.0155 | -2.7468 | -1.3684 |
| | (5.2983) | (2.4768) | (4.3147) | (3.6990) | (1.6968) |
| Log likelihood | -37.4968 | -35.2720 | -35.5830 | -37.4358 | -33.3714 |
| LR Chi-squared | | 4.4496** | 3.8276** | 0.1219 | 8.2507** |

| Number of observations | 33 | 33 | 33 | 33 | 33 |

**Panel E: Robustness test: determinants of the duration of short-term public debt contractions in AEs**

|  | (1) | (2) | (3) | (4) | (5) |
|---|---|---|---|---|---|
| Credit boom⊥ |  | 0.3099 |  |  | 0.4516 |
|  |  | (0.3203) |  |  | (0.2915) |
| House price boom⊥ |  |  | 0.2626 |  | 0.1915 |
|  |  |  | (0.2492) |  | (0.2385) |
| Equity price boom⊥ |  |  |  | 0.5299*** | 0.5530*** |
|  |  |  |  | (0.1609) | (0.1615) |
| Constant | 1.9959*** | 1.9988*** | 1.9933*** | 2.0048*** | 2.0054*** |
|  | (0.0905) | (0.0892) | (0.0908) | (0.0762) | (0.0733) |
| Weibull shape parameter (ln $p$) | 0.3136*** | 0.3180*** | 0.3225*** | 0.3192*** | 0.3375*** |
|  | (0.0762) | (0.0756) | (0.0759) | (0.0773) | (0.0750) |
| Frailty parameter (ln $\theta$) | -1.8814** | -1.9121** | -1.8081** | -3.2376 | -3.2861 |
|  | (0.8548) | (0.8578) | (0.8054) | (2.4296) | (2.4153) |
| Log likelihood | -169.4232 | -168.9236 | -168.8449 | -164.0175 | -162.4346 |
| LR Chi-squared |  | 0.9992 | 1.1565 | 10.8114*** | 13.9771*** |
| Number of observations | 136 | 136 | 136 | 136 | 136 |

**Panel F: Robustness test: determinants of the duration of short-term public debt contractions in EMs**

|  | (1) | (2) | (3) | (4) | (5) |
|---|---|---|---|---|---|
| Credit boom⊥ |  | -0.9804*** |  |  | -0.9023*** |
|  |  | (0.3460) |  |  | (0.3315) |
| House price boom⊥ |  |  | -0.3152 |  | -0.3892* |
|  |  |  | (0.2597) |  | (0.2405) |
| Equity price boom⊥ |  |  |  | 0.4909* | 0.4665* |
|  |  |  |  | (0.2646) | (0.2442) |
| Constant | 1.9418*** | 1.9203*** | 1.9628*** | 1.9407*** | 1.9452*** |
|  | (0.1228) | (0.1180) | (0.1019) | (0.1156) | (0.0920) |
| Weibull shape parameter (ln $p$) | 0.2576** | 0.3251*** | 0.2426*** | 0.2820*** | 0.3385*** |
|  | (0.1128) | (0.0999) | (0.0891) | (0.1061) | (0.0885) |
| Frailty parameter (ln $\theta$) | -2.9561 | -2.4501 | -14.7903 | -3.4157 | -14.4996 |
|  | (2.6070) | (1.5796) | (1216.6224) | (3.5173) | (887.0181) |
| Log likelihood | -85.1205 | -82.3887 | -84.5518 | -83.2672 | -79.8611 |
| LR Chi-squared |  | 5.4637** | 1.1374 | 3.7066** | 10.5189** |
| Number of observations | 67 | 67 | 67 | 67 | 67 |

**Panel G: Robustness test: determinants of the duration of medium-term public debt contractions in AEs**

|  | (1) | (2) | (3) | (4) | (5) |
|---|---|---|---|---|---|
| Credit boom$^\perp$ |  | -0.2274 |  |  | -0.0241 |
|  |  | (0.4212) |  |  | (0.4225) |
| House price boom$^\perp$ |  |  | 0.4153* |  | 0.3707 |
|  |  |  | (0.2202) |  | (0.2513) |
| Equity price boom$^\perp$ |  |  |  | 0.2117 | 0.0781 |
|  |  |  |  | (0.1817) | (0.1950) |
| Constant | 2.4966*** | 2.4887*** | 2.4863*** | 2.4999*** | 2.4893*** |
|  | (0.1153) | (0.1172) | (0.1146) | (0.1134) | (0.1141) |
| Weibull shape parameter (ln $p$) | 0.5874*** | 0.6006*** | 0.6426*** | 0.5924*** | 0.6375*** |
|  | (0.1170) | (0.1179) | (0.1168) | (0.1161) | (0.1178) |
| Frailty parameter (ln $\theta$) | -0.8737 | -0.7731 | -0.6575 | -0.9471 | -0.7158 |
|  | (0.8406) | (0.8209) | (0.7942) | (0.8634) | (0.8214) |
| Log likelihood | -73.2048 | -73.0765 | -71.4083 | -72.5108 | -71.3275 |
| LR Chi-squared |  | 0.2564 | 3.5929** | 1.3879 | 3.7545 |
| Number of observations | 69 | 69 | 69 | 69 | 69 |

**Panel H: Robustness test: determinants of the duration of medium-term public debt contractions in EMs**

|  | (1) | (2) | (3) | (4) | (5) |
|---|---|---|---|---|---|
| Credit boom$^\perp$ |  | -1.4717** |  |  | -1.4528*** |
|  |  | (0.5868) |  |  | (0.4999) |
| House price boom$^\perp$ |  |  | 0.4098 |  | 0.4624 |
|  |  |  | (0.3398) |  | (0.3055) |
| Equity price boom$^\perp$ |  |  |  | 0.1176 | 0.0115 |
|  |  |  |  | (0.3254) | (0.3009) |
| Constant | 2.7127*** | 2.6462*** | 2.5620*** | 2.7047*** | 2.3719*** |
|  | (0.1852) | (0.2429) | (0.2947) | (0.1886) | (0.3460) |
| Weibull shape parameter (ln $p$) | 0.4895** | 0.5976*** | 0.6361*** | 0.4995** | 0.8193*** |
|  | (0.2057) | (0.2310) | (0.2194) | (0.2068) | (0.1786) |
| Frailty parameter (ln $\theta$) | -1.6307 | -0.9484 | -0.0472 | -1.4977 | 1.0353 |
|  | (2.3447) | (2.2591) | (1.7330) | (2.1936) | (1.5001) |
| Log likelihood | -32.6226 | -30.9027 | -32.0427 | -32.5564 | -29.9023 |
| LR Chi-squared |  | 3.4397 | 1.1598 | 0.1324 | 5.4406 |
| p for model test |  | 0.0636 | 0.2815 | 0.7160 | 0.1422 |
| Number of observations | 29 | 29 | 29 | 29 | 29 |

Note: This table reports the orthogonal robustness tests for the impact of financial cycles on the duration of public

debt cycles. All regressions control the shared frailty. Standard errors are in parentheses. $p$ is the Weibull shape parameter, reflecting duration time-dependency. $\theta$ is the variance of $\alpha_i$ in Eq. (4). Credit boom$^\perp$ (House price boom$^\perp$ and Equity price boom$^\perp$) is the residual of the OLS linear regression in which Credit boom$^\perp$ (House price boom$^\perp$ and Equity price boom$^\perp$) is the dependent variable, and Credit growth rate and House price growth rate are the independent variables. Credit bust$^\perp$ (House price bust$^\perp$ and Equity price bust$^\perp$) is defined similarly. *, ** and *** represent significance at the 10%, 5% and 1% level, respectively.

Table D.3 presents the robustness tests for the impact of financial cycles on the amplitude of public debt cycles by including macroeconomic variables. Regression (1) to (6) are the results of regressions with different macroeconomic variables added one by one, on the basis of the three dummy variables and growth rate variables.

**Table D.3. Robustness test (with macroeconomic variables): determinants of the amplitude of public debt cycles.**

**Panel A: Robustness test: determinants of the amplitude of short-term public debt expansions in AEs**

|  | (1) | (2) | (3) | (4) | (5) | (6) |
|---|---|---|---|---|---|---|
| Credit bust | 1.0673 | -7.1536 | -0.2328 | -0.0048 | -5.1353 | -0.8554 |
|  | (16.0970) | (15.1761) | (16.1315) | (16.2823) | (15.7256) | (16.1955) |
| House price bust | -20.1088* | -20.8747** | -19.9848* | -19.7250* | -25.2968** | -19.3992* |
|  | (10.5574) | (9.7687) | (10.4365) | (10.4574) | (10.3060) | (10.4155) |
| Equity price bust | 12.7513* | 11.2037 | 12.6919* | 12.2429 | 13.1929* | 12.1332 |
|  | (7.5759) | (7.0817) | (7.5501) | (7.6124) | (7.3260) | (7.5644) |
| Credit growth rate | 7.3359** | 6.2954** | 6.9822** | 7.2933** | 6.6964** | 7.2623** |
|  | (2.8322) | (2.6524) | (2.8399) | (2.8270) | (2.7424) | (2.8189) |
| House price growth rate | -5.1789** | -5.8995*** | -5.1898** | -5.2607** | -3.4963* | -4.9271** |
|  | (2.0200) | (1.8747) | (1.9963) | (2.0595) | (2.0028) | (1.9928) |
| GDP | 1.3437 |  |  |  |  |  |
|  | (3.3552) |  |  |  |  |  |
| Monetary |  | 10.1195*** |  |  |  |  |
|  |  | (2.4137) |  |  |  |  |
| CPI |  |  | 6.0978 |  |  |  |
|  |  |  | (7.3794) |  |  |  |
| REER |  |  |  | 1.0282 |  |  |
|  |  |  |  | (2.5008) |  |  |
| Balance |  |  |  |  | -3.2040*** |  |
|  |  |  |  |  | (1.1113) |  |
| Oil |  |  |  |  |  | 0.3902 |
|  |  |  |  |  |  | (0.4448) |
| Constant | 32.0888*** | 19.4258*** | 29.5078*** | 32.5319*** | 40.3084*** | 30.8409*** |
|  | (4.5811) | (5.2504) | (5.6292) | (4.6202) | (5.2309) | (4.8253) |
| Log likelihood | -761.8486 | -751.5738 | -761.5194 | -761.8431 | -756.8675 | -761.4648 |
| Number of observations | 153 | 153 | 153 | 153 | 153 | 153 |

**Panel B: Robustness test: determinants of the amplitude of short-term public debt expansions in EMs**

|  | (1) | (2) | (3) | (4) | (5) | (6) |
|---|---|---|---|---|---|---|
| Credit bust | -5.9788 | -8.2883 | -7.8346 | -8.1249 | -6.1916 | -7.6330 |
|  | (21.3124) | (21.1902) | (21.2343) | (22.0048) | (21.4514) | (21.4335) |
| House price bust | 27.8467** | 27.4573** | 26.7202** | 27.2874** | 27.7675** | 26.8302** |
|  | (12.8310) | (12.7737) | (12.8166) | (12.9300) | (12.9019) | (13.0480) |
| Equity price bust | 11.1915 | 10.9413 | 10.0413 | 12.3665 | 14.4988 | 12.8366 |
|  | (11.4445) | (11.4097) | (11.6247) | (11.4915) | (11.8588) | (11.5603) |
| Credit growth rate | 1.6527 | 1.7034 | 2.6905 | 1.0706 | 1.3522 | 1.2425 |
|  | (3.9505) | (3.9275) | (4.2584) | (3.9678) | (3.9458) | (4.0582) |
| House price growth rate | -7.3008*** | -7.6208*** | -7.9545*** | -7.3031*** | -7.1389*** | -7.3539*** |
|  | (2.4334) | (2.4392) | (2.5138) | (2.4671) | (2.4642) | (2.4549) |
| GDP | 3.1074 |  |  |  |  |  |
|  | (3.3484) |  |  |  |  |  |
| Monetary |  | 1.8780 |  |  |  |  |
|  |  | (1.7220) |  |  |  |  |
| CPI |  |  | 2.4227 |  |  |  |
|  |  |  | (2.4967) |  |  |  |
| REER |  |  |  | -0.2646 |  |  |
|  |  |  |  | (2.1170) |  |  |
| Balance |  |  |  |  | 0.8825 |  |
|  |  |  |  |  | (1.3987) |  |
| Oil |  |  |  |  |  | 0.1256 |
|  |  |  |  |  |  | (0.5337) |
| Constant | 48.5571*** | 47.8920*** | 50.7146*** | 51.5237*** | 50.4135*** | 51.3131*** |
|  | (7.9831) | (8.0223) | (7.4149) | (7.5759) | (7.5755) | (7.4554) |
| Log likelihood | -345.7769 | -345.5358 | -345.7177 | -346.4053 | -346.1197 | -346.3755 |
| Number of observations | 72 | 72 | 72 | 72 | 72 | 72 |

**Panel C: Robustness test: determinants of the amplitude of medium-term public debt expansions in AEs**

|  | (1) | (2) | (3) | (4) | (5) | (6) |
|---|---|---|---|---|---|---|
| Credit bust | 45.3222 | 46.6692 | 44.4113 | 45.8752 | 38.4687 | 45.8416 |
|  | (31.0079) | (28.2168) | (30.9147) | (30.9685) | (31.7660) | (30.9415) |
| House price bust | -17.9347 | -15.1901 | -21.2011 | -17.4314 | -21.9194 | -17.6774 |
|  | (21.9869) | (19.9744) | (22.7675) | (22.1774) | (22.1581) | (21.9522) |
| Equity price bust | -5.1986 | -3.4039 | -3.5506 | -5.6827 | -5.3287 | -5.5719 |
|  | (16.8813) | (15.4222) | (17.2061) | (17.0805) | (16.7608) | (16.9704) |
| Credit growth rate | 20.8108** | 12.4290 | 21.5040** | 20.6022** | 18.0287* | 20.6528** |
| | (8.5706) | (8.1817) | (8.6344) | (8.5372) | (9.0204) | (8.5114) |
| House price growth rate | -5.3218 | -8.5974* | -5.5856 | -5.8444 | -3.8141 | -5.2067 |
| | (4.7559) | (4.3222) | (4.6134) | (5.2806) | (4.9361) | (4.7130) |
| GDP | -0.5469 | | | | | |
| | (8.5007) | | | | | |
| Monetary | | 21.7336*** | | | | |
| | | (6.7522) | | | | |
| CPI | | | -7.5922 | | | |
| | | | (15.9978) | | | |
| REER | | | | -0.0199 | | |
| | | | | (0.1152) | | |
| Balance | | | | | -1.8784 | |
| | | | | | (2.1995) | |
| Oil | | | | | | 0.1851 |
| | | | | | | (0.9721) |
| Constant | 61.7483*** | 25.7558* | 65.6778*** | 62.1319*** | 65.0614*** | 60.6758*** |
| | (10.3890) | (14.6036) | (13.3160) | (10.6557) | (10.9888) | (11.5599) |
| Log likelihood | -440.9642 | -433.3366 | -440.7860 | -440.9434 | -440.3825 | -440.9382 |
| Number of observations | 84 | 84 | 84 | 84 | 84 | 84 |

**Panel D: Robustness test: determinants of the amplitude of medium-term public debt expansions in EMs**

| | (1) | (2) | (3) | (4) | (5) | (6) |
|---|---|---|---|---|---|---|
| Credit bust | 60.4337* | 43.1803 | 43.4176 | 45.5607 | 45.1520 | 43.9177 |
| | (30.7302) | (33.7083) | (33.5788) | (30.0511) | (32.3809) | (34.2816) |
| House price bust | 45.3567* | 40.1920 | 40.5148 | 36.8252 | 39.4635 | 41.0898 |
| | (21.7932) | (25.4258) | (25.5448) | (22.2987) | (23.9087) | (26.3264) |
| Equity price bust | -10.8009 | 34.8802 | 34.4423 | 12.0751 | 37.8228 | 34.6578 |
| | (38.2795) | (34.4626) | (33.2236) | (32.9123) | (32.3668) | (33.3272) |
| Credit growth rate | 7.4051 | 8.8797 | 8.7593 | 7.8047 | 15.3255 | 8.6406 |
| | (11.9292) | (13.8750) | (13.7236) | (12.2264) | (15.2125) | (13.6731) |
| House price growth rate | -10.3936* | -11.9752* | -11.8686* | -12.8359** | -12.7507** | -11.9489* |
| | (4.9539) | (6.0853) | (5.6782) | (5.0495) | (5.4947) | (5.7231) |
| GDP | 21.1523* | | | | | |
| | (11.6261) | | | | | |
| Monetary | | -0.1781 | | | | |
| | | (3.8558) | | | | |
| CPI | | | -0.0496 | | | |
| | | | (8.7099) | | | |
| REER | | | | 6.7037 | | |
| Balance |  |  |  |  | 2.2528 |  |
|  |  |  |  | (4.1560) | (2.6218) |  |
| Oil |  |  |  |  |  | -0.0866 |
|  |  |  |  |  |  | (1.3013) |
| Constant | 85.8870*** | 97.5311*** | 96.9804*** | 103.2868*** | 100.1033*** | 96.8972*** |
|  | (13.7326) | (19.2910) | (17.5334) | (13.2396) | (14.0992) | (14.0546) |
| Log likelihood | -148.7843 | -153.1218 | -153.1249 | -149.6218 | -152.0531 | -153.1183 |
| Number of observations | 33 | 33 | 33 | 33 | 33 | 33 |

**Panel E: Robustness test: determinants of the amplitude of short-term public debt contractions in AEs**

|  | (1) | (2) | (3) | (4) | (5) | (6) |
|---|---|---|---|---|---|---|
| Credit boom | -2.3581 | -2.4162 | -2.4062 | -3.0014 | -2.0630 | -2.4408 |
|  | (4.7122) | (4.7943) | (4.7726) | (4.8559) | (4.7667) | (4.7744) |
| House price boom | 0.1020 | 0.6464 | 0.5519 | 0.6458 | 1.6251 | 0.5534 |
|  | (3.4995) | (3.5310) | (3.5461) | (3.5235) | (3.6539) | (3.5692) |
| Equity price boom | -5.5607** | -5.3386** | -5.3420** | -5.5176** | -5.1686** | -5.3890** |
|  | (2.3072) | (2.3343) | (2.3328) | (2.3472) | (2.3295) | (2.3510) |
| Credit growth rate | -0.9654 | -0.7611 | -0.7579 | -0.7348 | -0.7762 | -0.7583 |
|  | (0.6764) | (0.6739) | (0.6738) | (0.6740) | (0.6711) | (0.6740) |
| House price growth rate | -0.9433 | -0.7952 | -0.8157 | -0.9311 | -0.7008 | -0.7999 |
|  | (0.7328) | (0.7494) | (0.7411) | (0.7683) | (0.7396) | (0.7376) |
| GDP | 1.9282* |  |  |  |  |  |
|  | (1.1572) |  |  |  |  |  |
| Monetary |  | 0.0089 |  |  |  |  |
|  |  | (0.6770) |  |  |  |  |
| CPI |  |  | 0.6351 |  |  |  |
|  |  |  | (2.3301) |  |  |  |
| REER |  |  |  | 0.3806 |  |  |
|  |  |  |  | (0.6135) |  |  |
| Balance |  |  |  |  | -0.3409 |  |
|  |  |  |  |  | (0.3492) |  |
| Oil |  |  |  |  |  | -0.0161 |
|  |  |  |  |  |  | (0.0911) |
| Constant | -8.2093*** | -7.2528*** | -7.5247*** | -7.0226*** | -6.5219*** | -7.1952*** |
|  | (1.5238) | (1.6878) | (1.7661) | (1.4673) | (1.6003) | (1.4500) |
| Log likelihood | -491.7908 | -493.5655 | -493.5175 | -493.3168 | -492.9512 | -493.5453 |
| Number of | 136 | 136 | 136 | 136 | 136 | 136 |

observations

**Panel F: Robustness test: determinants of the amplitude of short-term public debt contractions in EMs**

|  | (1) | (2) | (3) | (4) | (5) | (6) |
|---|---|---|---|---|---|---|
| Credit boom | 6.2326 | 7.1493* | 6.4505 | 6.4682* | 6.3237 | 6.1220 |
|  | (3.9958) | (3.7133) | (3.9725) | (3.9214) | (3.8900) | (3.9258) |
| House price boom | -2.8398 | -2.5448 | -2.8965 | -2.6508 | -2.9950 | -2.6717 |
|  | (3.1337) | (2.8782) | (3.1261) | (3.0485) | (3.0402) | (3.0398) |
| Equity price boom | -2.8386 | -3.2203 | -2.4220 | -1.6789 | -2.7662 | -2.9039 |
|  | (2.9760) | (2.6158) | (2.8036) | (3.0469) | (2.7408) | (2.7820) |
| Credit growth rate | 0.1454 | -0.4852 | 0.2246 | 0.0897 | 0.0230 | 0.1152 |
|  | (0.3514) | (0.3986) | (0.3916) | (0.3330) | (0.3390) | (0.3300) |
| House price growth rate | -0.1896 | -0.1715 | -0.2412 | -0.2910 | -0.3159 | -0.2046 |
|  | (0.4871) | (0.4375) | (0.4767) | (0.4720) | (0.4660) | (0.4624) |
| GDP | -0.1416 |  |  |  |  |  |
|  | (0.5167) |  |  |  |  |  |
| Monetary |  | -0.9266** |  |  |  |  |
|  |  | (0.3826) |  |  |  |  |
| CPI |  |  | -0.5596 |  |  |  |
|  |  |  | (1.0555) |  |  |  |
| REER |  |  |  | 0.2318 |  |  |
|  |  |  |  | (0.3513) |  |  |
| Balance |  |  |  |  | -0.3571 |  |
|  |  |  |  |  | (0.3312) |  |
| Oil |  |  |  |  |  | -0.0869 |
|  |  |  |  |  |  | (0.1045) |
| Constant | -8.3375*** | -6.3399*** | -7.9129*** | -8.6303*** | -8.2805*** | -8.0260*** |
|  | (1.6962) | (1.5839) | (1.7350) | (1.3504) | (1.3752) | (1.5222) |
| Log likelihood | -212.1739 | -208.0393 | -209.3284 | -211.9012 | -211.3575 | -211.7083 |
| Number of observations | 67 | 67 | 67 | 67 | 67 | 67 |

**Panel G: Robustness test: determinants of the amplitude of medium-term public debt contractions in AEs**

|  | (1) | (2) | (3) | (4) | (5) | (6) |
|---|---|---|---|---|---|---|
| Credit boom | 11.4983 | 9.5782 | 11.3934 | 11.7109 | 11.9435 | 12.9353 |
|  | (11.0582) | (10.9836) | (11.0291) | (10.8403) | (10.8655) | (10.7956) |
| House price boom | -6.0684 (7.1962) | -7.9639 (7.1941) | -6.2000 (7.1230) | -5.4921 (6.9548) | -3.4052 (7.2634) | -5.8782 (6.8990) |
| Equity price boom | -3.4199 (5.2497) | -2.6620 (5.1763) | -2.5489 (5.6280) | -4.9934 (5.2509) | -4.8383 (5.2468) | -4.3314 (5.1067) |
| Credit growth rate | -2.0895 (2.1882) | -1.9501 (2.1411) | -2.2216 (2.2067) | -1.6447 (2.1631) | -1.9936 (2.1432) | -2.0043 (2.1224) |
| House price growth rate | 1.6310 (1.5562) | 2.1043 (1.5608) | 1.5473 (1.5651) | 2.0197 (1.5441) | 1.6689 (1.5206) | 1.8683 (1.5121) |
| GDP | 0.5520 (2.3721) | | | | | |
| Monetary | | -1.4676 (1.2283) | | | | |
| CPI | | | -2.0538 (4.3466) | | | |
| REER | | | | 0.4713 (0.3878) | | |
| Balance | | | | | -0.6614 (0.5764) | |
| Oil | | | | | | 0.2415 (0.1674) |
| Constant | -13.1652*** (3.3289) | -10.8380*** (3.4086) | -11.8085*** (3.7221) | -11.9209*** (3.0613) | -11.0863*** (3.3424) | -13.0978*** (2.9495) |
| Log likelihood | -248.3807 | -247.1575 | -248.2278 | -247.1141 | -247.2548 | -246.5909 |
| Number of observations | 69 | 69 | 69 | 69 | 69 | 69 |

**Panel H: Robustness test: determinants of the amplitude of medium-term public debt contractions in EMs**

| | (1) | (2) | (3) | (4) | (5) | (6) |
|---|---|---|---|---|---|---|
| Credit boom | 15.6439 (8.9666) | 13.0555 (8.6865) | 14.0476 (8.6968) | 21.6190** (9.0501) | 15.3787 (9.1587) | 13.1723** (5.0204) |
| House price boom | -12.7138* (6.3330) | -12.3880* (5.9540) | -12.7757* (6.0852) | -12.6780* (5.5876) | -12.5137* (6.3663) | -15.5483*** (3.6306) |
| Equity price boom | 8.0839 (5.1201) | 5.6353 (5.5209) | 8.5260 (4.8539) | 9.7633* (4.5685) | 8.3312 (5.0586) | 7.2345** (2.8447) |
| Credit growth rate | -1.0325 (1.5896) | -1.6520 (1.2611) | -0.8013 (1.3783) | -0.0543 (1.4264) | -1.2722 (1.3150) | -1.0091 (0.7203) |
| House price | -0.1163 | -0.2007 | -0.0903 | -0.4456 | -0.0462 | 0.1037 |
| growth rate | (0.9861) | (0.9232) | (0.9320) | (0.8980) | (1.0372) | (0.5454) |
| GDP | 0.2297 (0.8630) | | | | | |
| Monetary | | -0.9674 (1.0064) | | | | |
| CPI | | | -1.4184 (1.8327) | | | |
| REER | | | | 0.8868 (0.6281) | | |
| Balance | | | | | 0.0312 (0.6869) | |
| Oil | | | | | | -0.3803*** (0.0972) |
| Constant | -12.2687** (4.0567) | -8.9527* (4.2310) | -10.7907** (3.1725) | -13.5111*** (3.2435) | -11.6884** (3.4289) | -9.5299*** (1.9608) |
| Log likelihood | -76.8165 | -75.1645 | -73.6505 | -73.3290 | -76.9583 | -60.1655 |
| Number of observations | 29 | 29 | 29 | 29 | 29 | 29 |

Note: This table reports the robustness test for the impact of financial cycles on the amplitude of public debt cycles. All regressions control the country's fixed effect. Standard errors are in parentheses. *, ** and *** represent significance at the 10%, 5% and 1% level, respectively.

Table D.4 shows the results of the orthogonal robustness tests for the impact of financial cycles on the amplitude of public debt cycles. (1) is the baseline result that only controls the fixed effect. (2) to (5) are the results including financial cycle variables.

Table D.4. Orthogonal robustness test: determinants of the amplitude of public debt cycles.

**Panel A: Robustness test: determinants of the amplitude of short-term public debt expansions in AEs**

|  | (1) | (2) | (3) | (4) | (5) |
|---|---|---|---|---|---|
| Credit bust$^\perp$ |  | 8.6169 |  |  | 6.1903 |
|  |  | (16.5070) |  |  | (16.4113) |
| House price bust$^\perp$ |  |  | -18.3346* |  | -20.1706* |
|  |  |  | (10.5489) |  | (10.7185) |
| Equity price bust$^\perp$ |  |  |  | 8.5490 | 10.8617 |
|  |  |  |  | (7.7687) | (7.8021) |
| Constant | 29.4963*** | 29.4963*** | 29.4963*** | 29.4963*** | 29.4963*** |
|  | (3.3349) | (3.3445) | (3.3087) | (3.3321) | (3.3078) |
| Log likelihood | -771.9569 | -771.7916 | -770.1445 | -771.2252 | -768.8816 |
| Number of observations | 153 | 153 | 153 | 153 | 153 |

**Panel B: Robustness test: determinants of the amplitude of short-term public debt expansions in EMs**

|  | (1) | (2) | (3) | (4) | (5) |
|---|---|---|---|---|---|
| Credit bust$^\perp$ |  | -11.1282 |  |  | -8.1557 |
|  |  | (23.6567) |  |  | (22.9883) |
| House price bust$^\perp$ |  |  | 25.2424* |  | 23.3587* |
|  |  |  | (13.5435) |  | (13.6002) |
| Equity price bust$^\perp$ |  |  |  | 19.1226 | 17.4201 |
|  |  |  |  | (12.3112) | (12.2157) |
| Constant | 52.5916*** | 52.5916*** | 52.5916*** | 52.5916*** | 52.5916*** |
|  | (4.7257) | (4.7602) | (4.6211) | (4.6651) | (4.6141) |
| Log likelihood | -357.5845 | -357.4345 | -355.2991 | -355.9819 | -353.8063 |
| Number of observations | 72 | 72 | 72 | 72 | 72 |

**Panel C: Robustness test: determinants of the amplitude of medium-term public debt expansions in AEs**

|  | (1) | (2) | (3) | (4) | (5) |
|---|---|---|---|---|---|
| Credit bust$^\perp$ |  | 32.3524 |  |  | 30.8614 |
|  |  | (30.9534) |  |  | (31.3762) |
| House price bust$^\perp$ |  |  | -22.6956 |  | -21.1383 |
|  |  |  | (21.7553) |  | (22.0511) |
| Equity price bust⊥ |  |  |  | -2.7313 | -2.7682 |
|  |  |  |  | (17.2423) | (17.3237) |
| Constant | 51.5526*** | 51.5526*** | 51.5526*** | 51.5526*** | 51.5526*** |
|  | (6.5521) | (6.5469) | (6.5471) | (6.6079) | (6.6068) |
| Log likelihood | -447.6317 | -446.8343 | -446.8373 | -447.6132 | -446.0990 |
| Number of observations | 84 | 84 | 84 | 84 | 84 |

**Panel D: Robustness test: determinants of the amplitude of medium-term public debt expansions in EMs**

|  | (1) | (2) | (3) | (4) | (5) |
|---|---|---|---|---|---|
| Credit bust⊥ |  | 56.5387 |  |  | 48.8628 |
|  |  | (32.8291) |  |  | (33.9942) |
| House price bust⊥ |  |  | 40.2983 |  | 38.9958 |
|  |  |  | (27.4125) |  | (26.5547) |
| Equity price bust⊥ |  |  |  | 30.1016 | 17.8457 |
|  |  |  |  | (33.0218) | (32.1950) |
| Constant | 106.6396*** | 106.6396*** | 106.6396*** | 106.6396*** | 106.6396*** |
|  | (8.7506) | (8.2847) | (8.4663) | (8.7945) | (8.1845) |
| Log likelihood | -165.1541 | -162.3481 | -163.0637 | -164.3187 | -159.7436 |
| Number of observations | 33 | 33 | 33 | 33 | 33 |

**Panel E: Robustness test: determinants of the amplitude of short-term public debt contractions in AEs**

|  | (1) | (2) | (3) | (4) | (5) |
|---|---|---|---|---|---|
| Credit boom⊥ |  | -1.0639 |  |  | -2.2886 |
|  |  | (4.7988) |  |  | (4.7590) |
| House price boom⊥ |  |  | -0.3413 |  | 0.5648 |
|  |  |  | (3.5480) |  | (3.5192) |
| Equity price boom⊥ |  |  |  | -5.1767** | -5.3448** |
|  |  |  |  | (2.2787) | (2.3268) |
| Constant | -9.7533*** | -9.7533*** | -9.7533*** | -9.7533*** | -9.7533*** |
|  | (0.8965) | (0.9004) | (0.9006) | (0.8802) | (0.8872) |
| Log likelihood | -498.3687 | -498.3383 | -498.3629 | -495.2507 | -495.0884 |
| Number of observations | 153 | 153 | 153 | 153 | 153 |

**Panel F: Robustness test: determinants of the amplitude of short-term public debt contractions in AEs**

|  | (1) | (2) | (3) | (4) | (5) |
|---|---|---|---|---|---|
| Credit boom⊥ |  | 6.9551* |  |  | 6.3404* |

|  | (1) | (2) | (3) | (4) | (5) |
|---|---|---|---|---|---|
|  |  | (3.7307) |  |  | (3.8041) |
| House price boom⊥ |  |  | -2.7216 |  | -2.7793 |
|  |  |  | (2.9826) |  | (2.9123) |
| Equity price boom⊥ |  |  |  | -3.3607 | -2.5219 |
|  |  |  |  | (2.6634) | (2.6664) |
| Constant | -9.0363*** | -9.0363*** | -9.0363*** | -9.0363*** | -9.0363*** |
|  | (0.8580) | (0.8375) | (0.8595) | (0.8530) | (0.8391) |
| Log likelihood | -215.8616 | -213.5660 | -215.2972 | -214.7904 | -212.2927 |
| Number of observations | 67 | 67 | 67 | 67 | 67 |

**Panel G: Robustness test: determinants of the amplitude of medium-term public debt contractions in AEs**

|  | (1) | (2) | (3) | (4) | (5) |
|---|---|---|---|---|---|
| Credit boom⊥ |  | 12.4096 |  |  | 9.6306 |
|  |  | (10.1737) |  |  | (10.9858) |
| House price boom⊥ |  |  | -9.6251* |  | -5.8610 |
|  |  |  | (5.6068) |  | (7.0822) |
| Equity price boom⊥ |  |  |  | -5.3029 | -3.2821 |
|  |  |  |  | (4.3875) | (5.2000) |
| Constant | -14.6508*** | -14.6508*** | -14.6508*** | -14.6508*** | -14.6508*** |
|  | (1.4511) | (1.4431) | (1.4200) | (1.4436) | (1.4377) |
| Log likelihood | -254.1519 | -252.9784 | -251.8650 | -252.9994 | -251.0743 |
| Number of observations | 69 | 69 | 69 | 69 | 69 |

**Panel H: Robustness test: determinants of the amplitude of medium-term public debt contractions in EMs**

|  | (1) | (2) | (3) | (4) | (5) |
|---|---|---|---|---|---|
| Credit boom⊥ |  | 12.4581 |  |  | 16.1236* |
|  |  | (9.7468) |  |  | (8.3372) |
| House price boom⊥ |  |  | -10.3593 |  | -14.5250** |
|  |  |  | (5.9230) |  | (5.6466) |
| Equity price boom⊥ |  |  |  | 2.1597 | 8.5436* |
|  |  |  |  | (5.3251) | (4.6735) |
| Constant | -14.2107*** | -14.2107*** | -14.2107*** | -14.2107*** | -14.2107*** |
|  | (1.5026) | (1.4673) | (1.3961) | (1.5534) | (1.2194) |
| Log likelihood | -90.1503 | -88.2995 | -86.8577 | -89.9529 | -80.2898 |
| Number of observations | 29 | 29 | 29 | 29 | 29 |

Note: This table reports the orthogonal robustness tests for the impact of financial cycles on the amplitude of public debt cycles. All regressions control the country's fixed effect. Standard errors are in parentheses. Credit

boom$^\perp$ (House price boom$^\perp$ and Equity price boom$^\perp$) is the residual of the OLS linear regression in which Credit boom$^\perp$ (House price boom$^\perp$ and Equity price boom$^\perp$) is the dependent variable, and Credit growth rate and House price growth rate are the independent variables. Credit bust$^\perp$ (House price bust$^\perp$ and Equity price bust$^\perp$) is defined similarly. *, ** and *** represent significance at the 10%, 5% and 1% level, respectively.